\documentclass[iop]{emulateapj}

\shorttitle{Planet Traps and the Origin of the Observed Mass-Period Relation}
\shortauthors{Hasegawa \& Pudritz}

\begin{document}

\title{Evolutionary Tracks of Trapped, Accreting Protoplanets: the Origin of the Observed Mass-Period Relation}

\author{Yasuhiro Hasegawa and Ralph E. Pudritz\altaffilmark{1}}
\affil{Department of Physics and Astronomy, McMaster University,
    Hamilton, ON L8S 4M1, Canada}
\email{YH:hasegay@physics.mcmaster.ca, REP:pudritz@physics.mcmaster.ca}

\altaffiltext{1}{Origins Institute, McMaster University, Hamilton, ON L8S 4M1, Canada}

\begin{abstract}
The large number of observed exoplanets ($\gtrsim $ 700) provides important constraints on their origin 
as deduced from the mass-period diagram of planets. The most surprising features in the diagram are 1) the (apparent) pile 
up of gas giants at a period of $\sim 500$ days ($\sim1$ AU) and 2) the so-called mass-period relation which indicates 
that planetary mass is an increasing function of orbital period. We construct the evolutionary tracks of growing 
planets at planet traps in evolving protoplanetary disks and show that they provide a good physical understanding of how these 
observational properties arise. The fundamental feature of our model is that inhomogeneities in protoplanetary disks give rise to 
multiple (up to 3) trapping sites for rapid (type I) planetary migration of planetary cores. The viscous evolution of disks results in 
the slow radial movement of the traps and their cores from large to small orbital periods. In our model, the slow inward motion 
of planet traps is coupled with the standard core accretion scenario for planetary growth. As planets grow, type II migration takes 
over.  Planet growth and radial movement are ultimately stalled by the dispersal of gas disks via photoevaporation. 
Our model makes a number of important predictions: that distinct sub-populations of planets that reflect the properties of planet 
traps where they have grown result in the mass-period relation; that the presence of these sub-populations naturally explains a 
pile-up of planets at $\sim 1$ AU; and that evolutionary tracks from the ice line do put planets at short periods and fill an earlier 
claimed "planet desert" - sparse population of planets in the mass-semi-major axis diagram.
\end{abstract}


\section{Introduction} \label{intro}

The availability of large samples of exoplanets is being used to constrain theories of planet formation in a statistical 
sense \citep{us07}. The standard theoretical tools for this are the so-called population synthesis models 
\citep{il04i,il08,mab09,il10}, wherein gas giants are considered to be formed by two main successive processes: 
the formation of cores by runaway \citep[e.g.][]{ws89} and oligarchic growth \citep[e.g.][]{ki98}, followed by 
gas accretion onto the cores \citep[e.g.][]{p96}. This mode of forming gas giants is referred to as the core-accretion 
scenario. The orbits of these accreting protoplanets are regulated by planetary migration that 
eventually determines the radial distribution of planets \citep{ward97}. The spirit of population synthesis models is 
to hypothesize that the diversity in the properties of observed exoplanets reflects the range of the (initial) disk 
environments in which planets are born. Fine tuning of the efficiency of various physical processes such as migration 
rates allows one to {\it qualitatively} reproduce the observations summarized in the mass-period diagram. 

\begin{figure}%[!ht]
\begin{center}
\includegraphics[width=9cm]{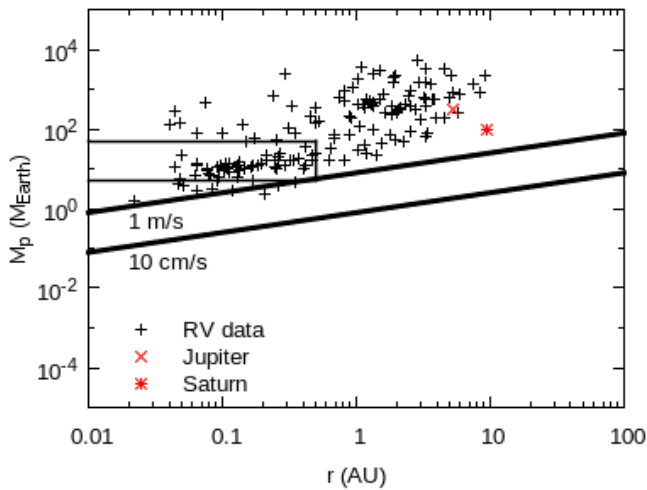}
\caption{Observed exoplanets using the radial velocity technique (denoted by black pluses). The data are obtained by 
the CORALIE and HARPS surveys, both of which are carried out through modest- and high-resolution spectrographs at La 
the Silla Observatory in Chile. We took the data from \citet{mml11}, wherein $\sim$ 150 observed exoplanets are selected 
from larger samples using consistent observational criteria for their statistical analyses. Thus, the data are well 
defined in order to discuss the statistical properties of exoplanets. Also, $M_p \sin i$ is plotted, since the 
inclinations $i$ are unknown. We have converted the data from orbital periods to semi-major axes for direct comparisons 
with our results (see Fig. \ref{fig5}). The host stars are F, G, or K stars. Jupiter (the red cross) and Saturn (the 
red star) are also shown for the reference. The thick black lines denote the amplitude of redial velocities of 
1 m s$^{-1}$ and 10 cm s $^{-1}$. The amplitude of 10 cm s $^{-1}$ is not achieved yet even in the HARPS survey while 
1 m s$^{-1}$ is well in hand. The data show the two trends; the (apparent) pile up of gas giants at $\sim 1$ AU  and 
the mass-period relation wherein planetary mass is an increasing function of period (particularly beyond 1 AU). Earlier 
papers \citep{il04i,il08v} predicted a planet desert demarcated by the black rectangle that covers in the range 
of planetary mass ($5 M_{\oplus} \lesssim M_p \lesssim 50 M_{\oplus}$) and of their semi-major axis 
(0.04 AU$ \lesssim r \lesssim $ 0.5 AU) in the diagram. The recent observations populate the desert.}
\label{fig1}
\end{center}
\end{figure}

Despite the success of these models, a single 
complete theory of planet formation that can reproduce the architecture of any (exo)planetary system including our 
Solar system is still unknown. In particular, it is unclear as to the physical origin of several key observations: the 
(apparent) pile-up of planets at $\sim1$ AU and the mass-period relation which shows that planetary mass increases with 
period (see Fig. \ref{fig1}).\footnote{Observations prefer orbital periods while semi-major 
axes are more natural in theoretical calculations. Since they are translatable through some analytical relations, we 
converted the observational data of \citet{mml11} from periods to semi-major axes using their published data of periods, 
planetary mass, eccentricities, and the amplitude of the radial velocities. Thus, we mainly use semi-major axes rather 
than periods.} Furthermore, there is a significant discrepancy between the theories and observations: the recent 
population synthesis models claimed that a planet desert - a region in the mass-period diagram with a lower population 
of exoplanets - is present in the range of 
planetary mass ($5 M_{\oplus} \lesssim M_p \lesssim 50 M_{\oplus}$) and of their semi-major axis 
(0.04 AU$ \lesssim r \lesssim $ 0.5 AU) \citep{il04i,il08v}\footnote{Recently, \citet{il10} succeeded 
in reproducing the population of low mass planets with short orbital radii by adding another physical process - mergers 
of protoplanets - that takes place after the gas disks are severely depleted. As shown below, on the contrary, our model 
is able to explain the population within the same framework of forming gas giants.} whereas many exoplanets are already 
observed there (see the black rectangle in Fig. \ref{fig1}).

In this paper, we address how inhomogeneities in protoplanetary disks can account for the observed trends. For this 
purpose, we constructed and followed evolutionary tracks of planets that grow at disk inhomogeneities. More 
specifically, we compute planetary growth and migration in protoplanetary disks that evolve with time due to disk 
viscosity and photoevaporation of gas, by tracking the movement of disk inhomogeneities such as dead zones, 
ice lines and heat transitions. The fundamental contribution of disk inhomogeneities to theories of planet formation 
here is that they give rise to trapping sites for rapid type I planetary migration of cores of gas giants 
\citep[often referred to as planet traps in the literature,][]{mmcf06,il08v,mpt09,hp10,lpm10}. Following the viscous 
evolution of disks, planet traps gradually move inwards in their disks, taking the trapped cores with them. The 
trapping and transport of cores is a central feature of our models in which protoplanets accrete gas as they move 
with the traps. We will show below that a semi-analytical model, wherein these two effects of planet traps, further 
planetary growth and subsequent type II migration that is terminated by photoevaporation of gas disks are all combined, 
can provide natural explanations of a number of the important observational properties: 1) the origin of the observed 
mass-period relation, 2) the origin of the pile up of observed gas giants at $\sim$ 1 AU, 3) the origin of low-mass 
planets distributing in the earlier claimed planet desert, 4) prediction of a new planet desert that originates from 
different physical processes than the earlier desert.     

The plan of this paper is the following. In $\S$ \ref{Ori_pt}, we summarize under what conditions planet traps are 
generated and which tidal torque and disk property play the most crucial role for activating the traps. In $\S$ \ref{diskmodels}, 
we describe disk models that are used for specifying the properties of disk inhomogeneities  while, in $\S$ 
\ref{disk_evol}, we discuss how these inhomogeneities evolve with time following viscous evolution of disks with 
photoevaporation of gas. In $\S$ \ref{charact_mass}, we derive the characteristic masses of planets that are captured 
at planet traps and how these masses define the mode of planetary migration. In $\S$ \ref{tracks}, we synthesize the 
above treatments and develop a semi-analytical model of planetary growth and migration affected by planet traps for 
constructing evolutionary tracks of growing planets in the mass-period diagram. We present our results and compare them 
with the observations in $\S$ \ref{resu}. The general reader may wish to skip to $\S$ \ref{resu} for a non-technical, 
astrophysical discussion of the results. Parameter studies are performed in $\S$ \ref{ParaStu}. $\S$ \ref{conc} is 
devoted to our discussion and conclusions. 

\section{Origins of planet traps} \label{Ori_pt}

Planet traps, a term first coined by \citet{mmcf06}, are one of the keys to resolving the long-standing problem of rapid type I 
migration that can lead to the loss of any planetary system to the host stars within $\sim 10^{5}$ years. This arises due 
to the high efficiency of angular momentum transfer between (proto)planets and their natal disks \citep{ward97,ttw02}. The 
basic idea of planet traps lies in the fact that the direction of migration can switch from inwards to outwards when 
planets migrate through the disks that have some kind of inhomogeneities. This is the combined consequence of the high 
sensitivity of type I migration to disk properties \citep{ttw02,pbck09,hp10c,hp11} and the density and temperature 
modifications produced by the disk inhomogeneities \citep{dccl98,mg04,mpt07,hp09b}. Since a complete discussion of how disk 
inhomogeneities give rise to planet traps is presented elsewhere \citep[e.g.][hereafter Paper I]{hp11}, we simply 
summarize the typical disk configurations with which planet traps are created and the responsible tidal torques and 
disk properties by which planet traps are activated with such disk configurations in Table \ref{table1}. In this paper, 
we focus on dead zones, ice lines, and heat transitions which all become planet traps (see $\S$ \ref{diskmodels} for 
their definitions, also see Paper I).

\begin{table*}
\begin{minipage}{15cm}
%\begin{table}
\begin{center}
\caption{Typical disk structures for generating planet traps}
\label{table1}
\begin{tabular}{ccc}
\hline
Torque                               &  Disk structures         &  Relevant disk properties   \\ \hline
Lindblad torque                      &  $t>1.5$ with $s=-1$     &  Dead zones                 \\
                                     &  $s<-2$  with $t=-0.5$   &  Dead zones and ice lines    \\
Vortensity-related corotation torque &  $s>1$ with $t\sim -0.5$ &  Inner edge of disks         \\
Entropy-related corotation torque    &  $t< -1.1$ with $s=-1$   &  Viscous heating             \\
\hline
\end{tabular}
\end{center}
Simple power-law disk structures are assumed, that is, the surface density of gas $\Sigma \propto r^{s}$ and 
the disk temperature $T \propto r^{t}$. We refer the reader to Paper I for a more complete discussion.
%\end{table}
\end{minipage}
\end{table*}

\section{Disk models} \label{diskmodels}

A significant number of analytical and numerical studies have shown that "realistic" disks are likely to possess 
several kinds of inhomogeneities \citep{g96,dccl98,mg04,mdk11}, which can activate planet traps. We first briefly 
describe our disk models that serve as the basis for specifying the properties of the disk inhomogeneities such as 
their positions and surface densities. We refer the reader to Paper I for the complete discussion.

We adopt the standard models of steady accretion disks that have accretion rates modeled as
\begin{equation}
 \dot{M}=3 \pi \nu \Sigma_g= 3 \pi \alpha c_s H \Sigma_g,
 \label{mdot}
\end{equation}
where $\Sigma_g$, $\nu=\alpha c_s H$, $c_s$, and $H$ are the surface density, the viscosity, the sound speed, and the 
pressure scale height of gas disks, respectively. The famous $\alpha-$prescription is assumed for characterizing the 
strength of disk turbulence \citep{ss73}. 

\subsection{Positions of disk inhomogeneities}

Adopting the standard disk model, we can estimate the positions of disk inhomogeneities. These positions are crucial 
because (proto)planets that undergo rapid type I migration will get trapped there.  We simply summarize 
the positions here and refer the reader to Paper I for the complete derivations (also see Table \ref{table1}). 

There are generally of three kinds of disk inhomogeneities: dead zones, ice lines and heat transitions (Paper I). Dead 
zones are present in the inner region of disks where high energy photons such as X-rays from the central stars and 
cosmic rays cannot penetrate \citep{g96,mp06,in06}. The defining feature of the dead zones is the low amplitude of 
turbulence there that results from the poor coupling of the magnetic field with weakly ionized disks \citep[so that 
magnetorotational instabilities (MRIs) are suppressed there, e.g.][]{b03}. Ice lines at which the disk temperature is 
low enough to trigger condensation of molecules such as water are the most famous and indispensable of disk 
inhomogeneities \citep{js04,mdk11}. They play an important role in population synthesis models \citep{il04i,mab09}. 
Heat transitions are also well recognized in the literature and arise at that radius at which the main heat source changes 
from viscous heating to stellar irradiation \citep[Paper I]{dccl98,mg04}. We have recently demonstrated that the heat 
transition becomes a planet trap based on analytical arguments \citep[Paper I, also see][]{kl12}. The validity of the heat 
transition traps has been recently confirmed by hydrodynamic simulations \citep{yi12}.  

In principle, the structure of dead zones can be specified by solving the ionization equations \citep{smun00,mp06,in06}. 
Nonetheless, the resultant structures depend sensitively on disk parameters that are difficult to determine through 
the observations. Therefore, we adopt a parameterized treatment of dead zones \citep[Paper I]{kl07,il08,mpt09} in which 
the effective $\alpha$ in the layered region can be given as 
\begin{equation}
 \alpha = \frac{\Sigma_A \alpha_A+(\Sigma_g-\Sigma_A) \alpha_D}{\Sigma_g},
 \label{mean_alpha}
\end{equation}
where $\alpha_A$ and $\alpha_D$ are the strength of turbulence in the active and dead layers, respectively, and 
the surface density of the active layer $\Sigma_A$ is modeled as 
\begin{equation}
 \Sigma_A=\Sigma_{A0} f_{ice}\left( \frac{r}{r_0} \right)^{s_A},
 \label{Sigma_A}
\end{equation}
where $r_0=1$ AU is the characteristic radius, $f_{ice}$ can contain the effects of ice lines. More specifically, 
$f_{ice}$ represents possible reductions of $\Sigma_A$ at the ice line that originate from the complex interplay 
between ice-coated, sticky dust grains and the absorption of free electrons by them \citep[Paper I, see a more complete 
discussion]{smun00,il08v}. Thus, the structures of dead zones are controlled only by $\Sigma_{A0}$ and $s_A$ in this 
formalism. This approach is useful because it enables one to investigate how important the structure of dead zones is 
for understanding the population of planets by simply varying parameters, $\Sigma_{A0}$ and $s_A$ (see $\S$ 
\ref{ParaStu}). Assuming stationary disk models (see equation (\ref{mdot})), the surface density of gas is given as
\begin{equation}
 \Sigma_g= \frac{\dot{M}}{3 \pi c_s H \alpha_D} - \Sigma_A \frac{\alpha_A - \alpha_D}{\alpha_D}.
\label{sigma_base}
\end{equation}

With equation (\ref{sigma_base}) in hand, the position of a dead zone trap is given as 
\begin{equation}
 \frac{r_{dz}}{r_0} = \left( \frac{\dot{M}}{3 \pi (\alpha_A+\alpha_D) \Sigma_{A0} H_0^2 \Omega_0} 
                         \right)^{\frac{1}{s_A+t+3/2}},
 \label{r_dz}
\end{equation}
where $H_0$ and $\Omega_0$ are the pressure scale height and Keplerian frequency at $r=r_0$. This can be derived from 
the assumption that the outer edge of dead zones is specified around $\Sigma_A \sim \Sigma_g/2$. 

The position of the ice line of molecular species $k$ is 
\begin{equation}
 \frac{r_{il}}{r_0} = \left[ \frac{1}{T_{m,k}^{12}(r_{il})}
                        \frac{27 \bar{\kappa}_0 \mu_g \Omega_0^3} {64 \sigma_{SB} \alpha_D \gamma k_B}
                        \left( \frac{\dot{M}}{3 \pi} \right)^2
                 \right]^{2/9} \propto \dot{M}^{4/9},
 \label{r_il}
\end{equation}
where $T_{m,k}$ is the disk midplane temperature below which molecules $k$ can condense, 
$\bar{\kappa}_0= 2\times 10^{16}$ is the opacity at the ice line of the molecule, $\mu_g$ is the mean molecular weight 
of the gas, $k_B$ is the Boltzmann constant, and $\gamma=1.4$ is the adiabatic index. This is given by the recent 
results which show that the viscous heating (rather than stellar irradiation) is generally dominant for determining 
the position of ice lines \citep[Paper I]{mdk11}. This expression is applicable for any molecules. Nonetheless, we 
focus on water ice lines here, since they are likely to be the most important molecule for understanding the observed 
mass-period relation (Paper I). For ice lines of water, $T_{m,\mbox{H}_2\mbox{O}}(r_{il})=170$ K \citep{js04}. 

For the case that ice lines are located within dead zones, the position of the trap needs to satisfy the following 
condition:
\begin{equation}
 \frac{r_{il}}{r_{dz}} > 
 \left( h(r_{dz}) \frac{\alpha_A + \alpha_D}{\alpha_A - \alpha_D} \right)^{\frac{1}{s_A+t/2+1}}.
 \label{condition_il}
\end{equation}

Finally, the position of heat transitions is written as
\begin{eqnarray}
 \label{r_ht}
 \frac{r_{ht}}{r_0} & = & \left[ \frac{1}{T_{m0}} \left( \frac{r_0}{R_*} \right)^{3/7} 
                          \left( \frac{27 \bar{\kappa}_0  \mu_g \Omega_0^3}{64 \sigma_{SB} \alpha_A \gamma k_B} 
                               \left( \frac{\dot{M}}{3 \pi} \right)^2  \right)^{1/3} 
                          \right]^{14/15} \\
                   & \propto & \dot{M}^{28/45}, \nonumber
\end{eqnarray}
where $\bar{\kappa}_0= 2 \times 10^{-4}$ is the opacity at the heat transitions, $R_*$ is stellar radius,  
\begin{equation}
 T_{m0} \simeq  \left( \frac{1}{H} \right)^{2/7} \left( \frac{T_*}{T_c} \right)^{1/7} T_*,
\end{equation}
\begin{equation}
 T_c \equiv \frac{GM_* \mu_g}{k_B R_*},
\end{equation}
$T_*$ and $M_*$ are stellar effective temperature and mass, respectively, and $G$ is the gravitational constant. We have 
adopted analytical models of \citet{cg97} for the temperature of the disk midplane heated by stellar irradiation. Planet 
traps arising from the heat transitions are active only if $r_{ht}>r_{dz}$.

\begin{figure}%[!ht]
\begin{center}
\includegraphics[width=9cm]{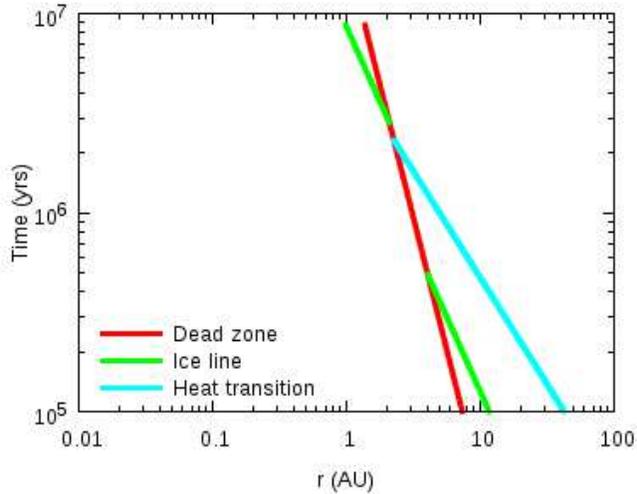}
\caption{Time evolution of the positions of three disk inhomogeneities for a disk around a classical T Tauri star 
(CTTS). The dead zone is denoted by red, the ice line by green, and the heat transition by light blue. Following the 
viscous 
evolution of disks, these zones move inwards at different rates. This results in the complex behaviors; convergence, 
disappearance, and re-appearance. Their end points are determined when photoevaporation of the gas disk takes place. }
\label{fig2}
\end{center}
\end{figure}

By comparing the positions of each disk inhomogeneity (see equations (\ref{r_dz}), (\ref{r_il}), and (\ref{r_ht})), 
one immediately observes that disk evolution, which lowers the accretion rate $\dot{M}$, moves them inwards, but at 
different rates (see Fig. \ref{fig2}). This is important for understanding the observed mass-period relation. 

\subsection{Characteristic surface densities at disk inhomogeneities}

We estimate the characteristic surface densities at the disk inhomogeneities, following the above formulation. 
Although the detailed structures of disk inhomogeneities remain to be simulated, a number of analytical and numerical 
studies based on the standard viscous disk theory clarified their characteristic structure \citep{mg04,mpt07,il08v}. 
These surface densities are utilized for deriving the characteristic masses of planets and following evolutionary tracks 
of planets that grow in planet traps. 

As briefly mentioned above, the outer edge of dead zones is determined by $\Sigma_A\sim \Sigma_g/2$. Substituting this 
condition into equation (\ref{sigma_base}), we find that the characteristic surface density at $r_{dz}$ is given as
\begin{equation}
 \Sigma_{g,dz} \approx \frac{2 \dot{M}}{3 \pi (\alpha_A+\alpha_D) r^2 h^2 \Omega},
 \label{sigma_dz}
\end{equation}
where $h=H/r$ is the aspect ratio.

At the ice lines, the surface density is approximately written as 
\begin{equation}
 \Sigma_{g,il} \approx \frac{\dot{M}}{3 \pi \alpha_D r^2 h^2 \Omega}.
 \label{sigma_il}
\end{equation}
We took the mean value of $\alpha$ as $\sim \alpha_D$. As mentioned before, this assumption is based on the recent 
extensive studies of ice lines \citep[Paper I]{smun00,il08v}. These studies indicate that ice lines can be regarded 
as a localized dead zone. 

On the other hand, the magnitude of turbulence at the heat transition is expected to be high enough to assume that disks 
are fully turbulent. As a result, the characteristic surface density at the heat transitions is given as 
\begin{equation}
 \Sigma_{g,ht} \approx \frac{\dot{M}}{3 \pi \alpha_A r^2 h^2 \Omega}.
 \label{sigma_ht}
\end{equation}

\section{Time evolution of disks and their inhomogeneities} \label{disk_evol}

Time evolution of protoplanetary disks is established by the combination of viscous turbulence and the photoevaporation 
of gas. These agents regulate the movement of disk inhomogeneities. We present our treatments of them.

\subsection{Viscous evolution}

Viscous turbulence is the dominant driver of disk evolution \citep[e.g.][]{a11}. We adopt similarity solutions 
\citep{lbp74} for constraining the relation between the accretion rate $\dot{M}$ and time $\tau$. Similarity solutions 
are derived from the conservation of angular momentum of disks \citep[also see][]{hcg98}. Considering a disk that has 
mass $M_{d}$ and a characteristic disk radius $R_c$, its angular momentum $J_{d}$ can be written as
\begin{equation}
 J_{d} \approx M_d R_c^{1/2}.
 \label{ang_mom}
\end{equation}
Following time evolution where disk material is accreted onto the central star, equation (\ref{ang_mom}) ensures that 
$R_c$ is an increasing function of time (since $J_{d}$ is roughly constant and $M_{d}$ steadily decreases with time). 
In a simplified analysis, the expansion rate of $R_c$ can be written as 
\begin{equation}
 \frac{d R_c}{dt} \approx \frac{R_c}{\tau_{vis}},
 \label{rateofR_c}
\end{equation}
where $\tau_{vis}=r^2/(3\nu)$ is the viscous timescale. Assuming a power-law structure for the disk temperature 
($T\propto r^{t}$), equation (\ref{rateofR_c}) gives
\begin{equation}
 R_d \propto \tau^{\frac{1}{(1/2-t)}},
\end{equation}
and the total disk mass $M_d$ decreases as (see equation (\ref{ang_mom}))
\begin{equation}
 M_d \propto \tau^{-1/(1-2t)}.
\end{equation}
As a result, the accretion rate is related to time through the following relation;
\begin{equation}
 \dot{M} \propto \tau^{-\frac{t-1}{t-1/2}}.
 \label{mdot_acc}
\end{equation}

Combining the observations which show that the median accretion rate for classical T Tauri stars (CTTSs) of age 
$\sim1$ Myrs is $\sim 10^{-8} M_{\odot}$ yr$^{-1}$ \citep{hcg98} and that $\dot{M} \propto M_*^2$ \citep{cmb04,mlb05}, 
we have the following scaling law for accretion rates;
\begin{equation}
 \dot{M} \simeq 10^{-8} M_{\odot} \mbox{ yr}^{-1} f_{acc} 
                                  \left( \frac{\tau}{ 10^{6} \mbox{ yr}} \right)^{-\frac{t-1}{t-1/2}} 
                                                   \left( \frac{M_*}{0.5M_{\odot}}\right)^2,
 \label{mdot_vis}
\end{equation}
where we have assumed that the typical mass of CTTSs is $\sim 0.5 M_{\odot}$ and introduced a dimensionless factor 
$f_{acc}$. This factor can be utilized for varying $\dot{M}$ and investigating the subsequent consequences on disk 
evolution and planet formation.

\subsection{Photoevaporation}

It is still unclear how gas disks disperse in the final stages of their evolution \citep[e.g.][]{a11}. One of the 
leading mechanisms is photoevaporation which arises from heating up gas by high energy photons from the surrounding stars 
and subsequent evaporation of gas due to the thermal pressure \citep{hjl94,jhb98}. In principle, photoevaporation rates are 
determined by the complex interplay between physical and chemical processes that take place in protoplanetary disks 
being irradiated by their central and nearby massive stars from far-UV (FUV) to extreme-UV (EUV) and up to X-rays 
\citep[references herein]{gh09}. Recent extensive studies have investigated how effective photoevaporation is in the 
dispersal of gas disks. In our models, we adopt a simple scaling law to represent the effects of photoevaporation. 

Following the treatment of \citet[see their Appendix for the complete derivation]{ahl04}, photoevaporation rates can be 
scaled as
\begin{equation}
 \dot{M}_{pe}=f_{pe} N_C \mu_g  c_s r_g \left( \frac{r_g}{r} \right) \exp \left( - \frac{r_g}{2r}\right),
 \label{mdot_pe}
\end{equation}
where $f_{pe}$ is a dimensionless factor of order unity, $N_C$ is the critical column density of gas that is heated by 
stellar radiation, and the gravitational radius $r_g$ is given as 
\begin{equation}
 r_g= \frac{GM_* \mu_g }{k_BT} \approx 100 \mbox{ AU} \left( \frac{T}{1000\mbox{ K}} \right)^{-1} 
                                                    \left( \frac{M_*}{1 M_{\odot}} \right).
 \label{r_g}
\end{equation}

Utilizing some of the most advanced results of photoevaporation, we further simplified equation (\ref{mdot_pe}). 
Recently, \citet{gh09} have investigated photoevaporation of gas disks by taking into account radiation of a central 
star that covers FUV, EUV and X-rays, and found that FUV heating plays the dominant role for inducing photoevaporation 
at $r\gtrsim $ 3 AU. This can be understood by the fact that photoevaporation rates are determined by the product of the 
gas temperature and density. EUV heating that leads to ionizing atomic hydrogen results in higher gas temperatures 
($\sim 10^{4}$ K) than FUV heating ($\sim 10^{2}-10^{3}$ K). Nonetheless, the ionization front above which EUV heating 
dominates can only penetrate the disk atmosphere where gas density is much lower than that where FUV heating becomes 
dominant. As a result, FUV-induced photoevaporation rates exceed EUV-induced ones. 

When photoevaporation is established mainly by FUV heating, the heated outgoing flow acts as an additional source of 
opacity for the FUV photons \citep{jhb98,ahl04}. Consequently, the critical column density heated up by FUV 
satisfies the self-regulation relation;
\begin{equation}
 \tau_{FUV} = \sigma_{FUV} N_C \sim 1,
 \label{tau_FUV}
\end{equation}
where $\sigma_{FUV} \approx 8 \times 10^{-22}$ cm$^{2}$ is the reasonable cross section of dust grains for the FUV 
photons. This enables us to specify $N_C$ in equation (\ref{mdot_pe}). Also, the peak of FUV-induced photoevaporation 
rates is attained around 0.1-0.2$r_g$, rather than $r_g$ that is valid for EUV-induced photoevaporation 
\citep{ahl04,gh09}. This can be again explained by the combination of the gas temperature and density, and is also 
confirmed by equation (\ref{mdot_pe}). 

Collecting the above arguments, we obtain a simplified, but physically motivated scaling law for photoevaporation rates;
\begin{equation}
 \dot{M}_{pe} \simeq 2.3 \times 10^{-9} M_{\odot} \mbox{ yr}^{-1} 
                              f_{pe} \left( \frac{c_s}{3 \mbox{ km s}^{-1}} \right)^{-1} 
                                      \left( \frac{M_*}{1 M_{\odot}} \right),
 \label{mdot_pescal}
\end{equation}
where we have used equations (\ref{r_g}) and (\ref{tau_FUV}) and set that $r=0.1r_g$ in equation (\ref{mdot_pe}). 

We note that equation (\ref{mdot_pescal}) can be applied for photoevaporation rates induced by both EUV and X-rays 
despite the fact that it is derived from the physical consideration based on FUV radiation. This can be 
done by adjusting the dimensionless factor $f_{pe}$ that is determined by the comparison with more detailed 
simulations. In fact, the conclusion of \citet{gh09} that FUV is the dominant source of photoevaporation is still a matter 
of debate in the literature. This is partly because they relied exclusively on hydro{\it static} solutions for quantifying 
winds driven by FUV radiation (although hydro{\it dynamical} models are needed for precisely estimating the winds), and 
partly because they adopted energy spectra which eventually reduce the effects of X-rays. As shown by 
\citet{oec10,oec11,oce12}, photoevaporation rates induced by X-rays can attain $\sim 10^{-8}$ M$_{\odot}$ yr$^{-1}$ 
for the most luminous X-ray sources. This high value is derived from employing observed Chandra spectra of TTSs and is 
comparable to the photoevaporation rate induced by FUV radiation. As a result, we intentionally avoid specifying the 
dominant source of photoevaporation of gas. Instead, we consider the general effects of photoevaporation on planet 
formation by treating $f_{pe}$ as a free parameter. 

\subsection{Photoevaporation of viscous disks}  

We are now in the position to discuss the complete treatment of disk evolution. We assume that the accretion rate 
through the disk is constant in space and regulated in time by equation (\ref{mdot_vis}). As time goes on, disk material 
accretes onto the central star and the accretion rate decreases. This change in $\dot{M}$ drives the movement of the 
disk inhomogeneities (see equations (\ref{r_dz}), (\ref{r_il}), and (\ref{r_ht})). When $\dot{M}$ becomes equal to the 
photoevaporation rate $\dot{M}_{pe}$, represented by equation (\ref{mdot_pescal}), we assume the gas disks to 
disperse completely. Although this treatment is somewhat idealized, it can account for the more detailed simulations. 
As an example, \citet{gdh09} investigated the evolution of viscous protoplanetary disks that are photoevaporated by the 
FUV, EUV, and X-ray radiation from their central star. They showed that viscous turbulence controls the early 
stage of disk evolution and the total disk mass gradually decreases initially, which can be formulated by power-laws. Once 
the condition that $\dot{M}\sim \dot{M}_{pe}$ is satisfied, the disk mass drops exponentially due to the combination of 
viscous evolution and photoevaporation of gas. They found that disks of initial mass $0.1 M_{\odot}$ around 
$\sim 1 M_{\odot}$ have the lifetime of $\sim 4 \times 10^{6}$ years.

We can derive similar results from our treatment by equating $\dot{M}$ with $\dot{M}_{pe}$: the disk lifetime is 
approximately estimated as $\sim 6 \times 10^{6}$ years. Thus, our treatment is sufficient for the purpose 
of representing the evolution of protoplanetary disks that are regulated by both viscosity and photoevaporation. 

In summary, we reduce the surface density of gas and the accretion rates, following equation (\ref{mdot_vis}). This 
results in the movement of the disk inhomogeneities. Also, we locate the final position of each disk inhomogeneity that 
is determined by the condition that $\dot{M} = \dot{M}_{pe}$.
 
\subsection{Parameters}

We summarize important parameters that establish the configuration and physical state of protoplanetary disks (see Table 
\ref{table2}). They are divided into three sets: stellar parameters ($M_*$, $T_*$, and $R_*$), the disk mass 
($\Sigma_{A0}$, $s_A$, $f_{acc}$, and $t$), and disk evolution ($\alpha_A$, $\alpha_D$, and $f_{pe}$). The disk 
evolution parameters can be translated into disk lifetimes. These three sets are fundamental to regulate planet 
formation in the disks, as confirmed in the population synthesis models \citep{il04i,mab09}. We focus on disks around 
CTTSs and denote the set of the values given in Table \ref{table2} as our fiducial model. Adopting these parameters, disk 
evolution proceeds from $\tau_{int}=10^{5}$ year to the time at $\dot{M}=\dot{M}_{pe}$ that defines the disk lifetime 
($\tau_{disk}$). A parameter study in which some of these quantities are changed is presented in $\S$ \ref{ParaStu}.

\begin{table*}
\begin{minipage}{15cm}
%\begin{table}
\begin{center}
\caption{Important disk quantities}
\label{table2}
\begin{tabular}{ccc}
\hline
Symbols        &  Meaning                                                                  & CTTSs           \\ \hline
$M_*$          &  Stellar mass                                                             & 0.5 $M_{\odot}$ \\
$R_*$          &  Stellar radius                                                           & 2.5 $R_{\odot}$ \\
$T_*$          &  Stellar effective temperature                                            & 4000 K          \\
$\Sigma_{A0}$  &  Surface density of active regions at $r=r_0$                             & 20  g cm$^{-2}$ \\           
$s_A$          &  Power-law index of $\Sigma_A (\propto r^{s_A})$                          & 3               \\ 
$t$            &  Power-law index of the disk temperature ($T \propto r^{t}$)              & -1/2            \\
$f_{acc}$      &  a dimensionless factor for $\dot{M}$ (see equation (\ref{mdot_vis}))     & 1               \\
$\alpha_{A}$   &  Strength of turbulence in the active zone                                & $10^{-3}$       \\        
$\alpha_{D}$   &  Strength of turbulence in the dead zone                                  & $10^{-4}$       \\   
$f_{pe}$       &  a dimensionless factor for $\dot{M}_{pe}$ (see equation (\ref{mdot_pe})) & 1/3             \\
\hline
\end{tabular}
\end{center}
%\end{table}
\end{minipage}
\end{table*}

\subsection{Movement of planet traps}

We draw upon our comprehensive 
analytical study (Paper I) which showed that the gas surface density and temperature modifications induced by the disk 
inhomogeneities are significant enough to reverse the direction of rapid type I migration. Therefore these positions 
are indeed trapping points of rapid type I migrators. 

Fig. \ref{fig2} shows the movement of all the three disk inhomogeneities. As demonstrated by Paper I, they all 
move inwards, but at different rates. The inward movements arise from the viscous evolution and the resultant reduction 
of the surface density of the disks (see equation (\ref{mdot_vis})). Also, different moving rates for the traps result 
in complex behaviors of the multiple inhomogeneities such as convergence (e.g. merging of the heat transition with the 
dead zone), disappearance, and re-emergence of inhomogeneities (e.g. the behavior of the ice line). When $\dot{M}$ 
equals $\dot{M}_{pe}$, photoevaporation quickly disperses gas in the disk and hence the movement of the planet traps is 
terminated. This also determines the lifetime of the disk which is $\sim 8.8 \times 10^{6}$ years in this configuration. 
The behavior of the planet traps is crucial for understanding the observed mass-period relation later.

\section{Characteristic masses} \label{charact_mass}

We describe four characteristic masses that are important in our models (see Table \ref{table3}). Using the positions 
and characteristic surface densities at disk inhomogeneities given in $\S$ \ref{diskmodels}, these masses define the 
mode of planetary migration (see $\S$ \ref{tracks}). Also, they result in a segment of the mass-semi-major 
axis diagram (see Appendix \ref{app1}). 

\begin{table}
\begin{center}
\caption{Characteristic masses}
\label{table3}
\begin{tabular}{ccc}
\hline
Symbols      &  Meaning                             &  Equation         \\ \hline
$M_{mig,I}$  &  Minimum mass of type I migrators    &  (\ref{M_typeI})  \\
$M_{gap}$    &  Gap-opening mass                    &  (\ref{M_gap})    \\
$M_{crit}$   &  Critical mass of type II migrators  &  (\ref{M_crit})   \\
$M_{max}$    &  Maximum mass of planets            &  (\ref{M_max})    \\
\hline
\end{tabular}
\end{center}
\end{table}

\subsection{Type I regime}

For planets of mass smaller than the gap-opening mass $M_{gap}$ (see below), type I migration is the main agent 
that governs their orbital distribution. In our models, rapid type I migration is halted at the planet traps. Hence the 
location of type I migrators is predicted by the positions of disk inhomogeneities (Paper I, see Fig. \ref{fig2}). 

It is important to define the minimum mass of planets that will be captured by the planet traps. As demonstrated 
numerically by \citet{lpm10}, planets captured in planet traps "drop-out" if the following condition is satisfied:
\begin{equation}
 \frac{\tau_{mig,I}}{\tau_{\nu}}>1,
 \label{condition_trap}
\end{equation}
where $\tau_{mig,I}$ is the timescale of type I planetary migration and $\tau_{\nu}$ is the timescale that determines 
the moving rates of planet traps. This relation expresses the fact that trapped planets drop-out if the speed of type I 
migration becomes less than that of the moving traps. In general, $\tau_{mig,I}$ is scaled as
\begin{equation}
 \tau_{mig,I}= \frac{M_p r_p^2 \Omega_p }{2 \Gamma},
\end{equation}
where 
\begin{equation}
 \Gamma= K_{mig} \left( \frac{M_p}{M_*} \right)^2 \frac{\Sigma_{g,p} r_p^4 \Omega_p^2}{h_p^2}
\end{equation}
with $K_{mig}=1-10$, depending on the optical thickness of the disk \citep{pbck09}. For stationary accretion disk models 
(see equation (\ref{mdot})), the minimum mass of type I migrators that can be captured at planet traps is given as
\begin{equation}
 M_{mig,I} = \frac{h_p^2 M_*^2}{2 K_{mig} \Sigma_{g,p} r_p^2 \Omega_p \tau_{\nu}}.
 \label{M_typeI}
\end{equation}
We set $\tau_{\nu}=10^{6}$ yrs, because the moving rates of planet traps are eventually regulated by disk lifetimes 
(see Fig. \ref{fig2}) and the observations revealed that the disk lifetime of any CTTS disk is an order of Myrs.

\subsection{Gap-opening mass}

The gap-opening mass $M_{gap}$ distinguishes type I migration from type II and is well discussed in the literature 
\citep[e.g.][]{ward97,mp06}. It arises when a planet becomes sufficiently massive that the torque it exerts on the disk 
opens a gap. There are two main arguments for estimating $M_{gap}$. The first one is the Hill radius analysis: the Hill 
radius should be larger than the pressure scale height for maintaining gap formation, otherwise gaps are closed by the 
gas pressure. The second argument arises from viscous disks. Disk viscosity that controls disk evolution plays the 
counteractive role for gap formation. Therefore, the tidal torque of a planet on their disks opens a gap if it 
exceeds the viscous torque.  Summarizing these arguments, $M_{gap}$ is given as \citep[e.g.][]{mp06}
\begin{equation}
 \frac{M_{gap}}{M_*} = \mbox{min}\left[ 3 h_p^3, \sqrt{40 \alpha h_p^5} \right].
 \label{M_gap}
\end{equation}

\subsection{Type II regime}

Planets of mass larger than $M_{gap}$ open up a gap in their disks and undergo so-called type II migration. In 
the type II regime, we define two characteristic masses. One of them is the critical mass ($M_{crit}$) above which the 
inertia of type II migrators is significant enough to prevent type II migration from proceeding as disks evolve 
(otherwise the timescale of type II migration is given as $\tau_{mig,II} \sim \tau_{vis}$). This effect is also known 
as a damming effect \citep{sc95,ipp99}. The critical mass $M_{crit}$ is defined by the local disk mass:
\begin{equation}
 M_{crit} = \pi \Sigma_{g,p} r_p^2,
 \label{M_crit}
\end{equation}
where $\Sigma_{g,p}$ is the surface density of gas disks at the position of a planet ($r=r_p$). 

The other characteristic mass is the maximum mass of planets. In general, gas accretion onto cores of gas giants is not 
fully terminated even if they form a gap in their disks \citep{lhdb09}. This suggests that the possible maximum mass of 
planets which start forming at time $\tau$  can be estimated as 
\begin{eqnarray}
 \label{M_max}
 M_{max}(\tau) & \simeq & \int^{\tau_{disk}}_{\tau} d\tau \dot{M}  \\ \nonumber
               &   =    & 5 \times 10^{-3} M_{\odot} 
                      f_{acc}\left( t - \frac{1}{2} \right)
                     \left( \frac{M_*}{0.5 M_{\odot}} \right)^{\frac{2t-1}{t-1}} \\ \nonumber
       & \times & \left[ \left( \frac{\dot{M}(\tau_{disk})}{10^{-8} f_{acc}M_{\odot} \mbox{ yr}^{-1}}  \right)^{-1/(2(t-1))} \right. \\ \nonumber
       &  &    \left. - \left( \frac{\dot{M}(\tau)}{10^{-8} f_{acc}M_{\odot} \mbox{ yr}^{-1}}  \right)^{-1/(2(t-1))}  
                    \right],
\end{eqnarray}
where equation (\ref{mdot_vis}) is used. 

In conclusion, the trapping regime is defined by $M_{gap}$ and $M_{mig,I}$ in which type I migrators follow the movement 
of the planet traps while the type II regime is defined by $M_{max}$ and $M_{gap}$, wherein the radial distribution of 
planets is established by the type II migration (see the bottom panel of Fig. \ref{figA} (Right) in Appendix \ref{app1}).

\section{Evolutionary tracks of growing planets in planet traps} \label{tracks}

Armed with the positions and surface density of planet traps ($\S$ \ref{diskmodels}) and four characteristic masses 
($\S$ \ref{charact_mass}), we now describe semi-analytical models of planetary growth and migration that are used for 
generating evolutionary tracks of accreting planets.

\subsection{Planetary growth}

The formation of gas giants is divided mainly into three stages \citep{ws89,ki98,p96}: formation of rocky cores 
through runaway and oligarchic growth (Stage I), the subsequent slow gas accretion of the cores and formation of 
envelopes surrounding them (Stage II), and collapse of the envelopes and runaway gas accretion onto their cores 
(Stage III). In order to model these three physical processes, we adopt the formulation of \citet{il04i} who first 
attempted to understand the statistics of the observed exoplanets by carrying out population synthesis analyses. In 
this formulation, these processes are treated by simple, analytical prescriptions that are derived from the detailed 
numerical simulations. In Appendix \ref{app2}, we briefly describe our treatments that slightly modify the original 
formulation, and refer the readers to \citet{il04i} for a complete discussion. We also present a parameter study 
in Appendix \ref{app3} for confirming the validity of our tiny modifications.

\subsection{Orbital evolution of planets}
   
Orbital evolution of planets is governed by planetary migration that arises from tidal interactions of the planets with 
the surrounding gaseous disks \citep{ward97,ttw02}. As discussed in $\S$ \ref{charact_mass}, the characteristic 
masses will classify planetary migration to four modes, depending on planetary mass: slower type I, trapped type I, the 
standard type II, and slower type II migration (see Table \ref{table3}). We discuss our treatments of them below.

{\it Slower type I migration}: This mode is applicable if planetary mass is smaller than $M_p < M_{mig,I}$ (see equation 
(\ref{M_typeI})). When planets satisfy this condition, the migration rate of these planets is much smaller than the 
moving rate of gas that is regulated by disk viscosity. This is the reason why we call this mode of migration the 
{\it slower} type I migration. Therefore, we assume that these planets remain in the same position with time.

{\it Trapped Type I migration}: When planets are in the trapping regimes, that is, $M_{mig,I} \leq M_p \leq M_{gap}$ 
(see equations (\ref{M_typeI}) and (\ref{M_gap})), the radial positions of these planets follow the movement of planet 
traps.

{\it The standard type II migration}: When the mass of planets in the range between $M_{gap} \leq M_p \leq M_{crit}$ 
(equations (\ref{M_gap}) and (\ref{M_crit})), they undergo type II migration that proceeds as the gas disks evolve; the 
type II migration timescale $\tau_{mig,II}$ equals $\tau_{vis}$. Therefore, the planets move inwards with the velocity 
written as
\begin{equation}
 v_{mig,II} \simeq - \frac{\nu}{r}.
\end{equation}

{\it Slower type II migration}: For planets with $M_p \gtrsim M_{crit}$, on the contrary, the type II migration rate 
slows down due to the inertia of the planets \citep{sc95,ipp99}. That is why we refer this mode as to {\it slower} 
type II migration. As a result, the velocity of the planets becomes
\begin{equation}
 v_{mig,slowII} \simeq - \frac{\nu}{r(1+f_{mig,slowerII} M_p/M_{crit})},
 \label{slower_typeII}
\end{equation}
where we have followed \citet{hn12} for taking into account the effects of the inertia of planets. In addition, 
we have introduced a new free parameter $f_{mig,slowerII}$. As shown below, both of the trapped type I and slower 
type II migration are important agents that regulate the radial distribution of planets in our model. Furthermore, it is 
currently uncertain how effective the inertia of planets is in slowing down the standard type II migration. Thus, it 
is useful to clarify the role of the slower type II migration by performing a parameter study wherein the value of 
$f_{mig,slowerII}$ varies (see $\S$ \ref{ParaStu3}). We set $f_{mig,slowerII}=1$ for our fiducial model. 
 
\subsection{Disk models}

We adopt the disk models discussed in $\S$ \ref{diskmodels}. More specifically, we use the characteristic surface 
densities at three disk inhomogeneities. In addition, the surface density of dust $\Sigma_d$ is required to examine 
planetary growth there. 

We simply assume that 
\begin{equation}
 \Sigma_d=f_{dtg} \Sigma_g,
\end{equation}
where $f_{dtg}$ is the dust-to-gas ratio. Table \ref{table4} summarizes the values of $f_{dtg}$ at each disk 
inhomogeneities. We took $f_{dtg}=0.01$ at the dead zone, because the dust mass is canonically about a hundredth of 
the gas mass in protoplanetary disks \citep[e.g.][]{dhkd07}. At the ice line, condensation of water increases the dust 
density there and beyond. Therefore, we used $f_{dtg}=0.05$ at the heat transition. The reason that $f_{dtg}=0.01$ at 
the ice line is that we have already taken into account the effect of the ice line on $\Sigma_g$ by reducing the mean 
value of $\alpha$ (see equation (\ref{sigma_il})). As a result, $f_{dtg}=0.01$ is reasonable for specifying $\Sigma_d$ 
there.

\begin{table}
\begin{center}
\caption{Values of $f_{dtg}$}
\label{table4}
\begin{tabular}{cccc}
\hline
           &  Dead zone  &  Ice line  &  Heat transition     \\ \hline
$f_{dtg}$  &  0.01       &  0.01      &  0.05                \\
\hline
\end{tabular}
\end{center}
\end{table}

\subsection{Initial conditions}

We choose a value for the initial mass of cores $\simeq 0.01M_{\oplus}$, which is sufficiently smaller than the mass 
that is finally obtained by the oligarchic growth \citep{ki98,ki02}. We confirmed that this choice does not affect 
our results. 

The cores start growing at a position $r$ at a time $\tau$. In principle, core formation takes place 
anywhere in disks. Nonetheless, we assume that the cores will quickly end up on one of the traps in the initial setup.
It is noted that the assumption does not always assure the cores to be initially captured at their traps. This is 
because trapping happens only if the mass of the cores is larger than $M_{mig,I}$ (see equation (\ref{M_typeI})). 
Although one may consider the assumption of the initial $\tau$ and $r$ to be somewhat artificial, this is not the case. 
As shown by \citet{il08}, planetary cores that undergo rapid type I migration do not contribute to the population of 
gas giants, (since they plunge into their central star within the disk lifetime). This implies that only the cores 
that experience slower type I migration will play an important role for reproducing the observed gas giants. Thus, 
it is reasonable to focus on planet formation proceeding only in planet traps that can substantially slow down the 
type I migration.  

Based on the assumption, it is only necessary to choose a distribution of the initial time $\tau$ (or position $r$) for 
the growth of cores to begin. The positions of disk inhomogeneities are related to the time $\tau$ through the 
accretion rate (see equation (\ref{mdot_vis})). Table \ref{table5} summarizes our 7 choices of $\tau$ which are selected 
to cover the entire disk lifetime within which planetary growth and migration take place. For reference purpose, the 
initial positions that are determined by equations (\ref{r_dz}), (\ref{r_il}), and (\ref{r_ht}) are also shown in the 
same table. It is noted that the multiple choices of the initial time result in forming multiple planets in each planet trap. 
We emphasize that the productivity of each planet trap - how many planets eventually form in each planet trap during 
the disk lifetime - and the relation between the number of finally formed planets and planet traps should be investigated 
separately. 

We neglect the planet-planet interactions of the cores that grow in different planet traps - we leave this for our 
future work.

\begin{table*}
\begin{minipage}{15cm}
%\begin{table}
\begin{center}
\caption{The initial times and positions}
\label{table5}
\begin{tabular}{cccc}
\hline
The initial time (yr)  &  Dead zone (AU) &  Ice line (AU)  &  Heat transition (AU)    \\ \hline
$10^{5}$               &  7.3            &  11.7           &  42.4                    \\
$2 \times 10^{5}$      &  5.7            &  7.4            &  22.3                    \\
$4 \times 10^{5}$      &  4.4            &  4.6            &  11.7                    \\
$8 \times 10^{5}$      &  3.4            &  N/A            &  6.1                     \\
$1.6 \times 10^{6}$    &  2.6            &  N/A            &  3.2                     \\
$3.2 \times 10^{6}$    &  2.0            &  1.9            &  N/A                     \\
$6.4 \times 10^{6}$    &  1.6            &  1.2            &  N/A                     \\

\hline
\end{tabular}
\end{center}
We assume that planet formation does not take place in a planet trap when the planet trap disappears due to convergence 
with a dead zone trap. N/A represents such cases. 
%\end{table}
\end{minipage}
\end{table*}

\subsection{Concurrent evolution of planetary growth and migration}

We may now follow the evolutionary tracks in the mass-semi-major axis diagram for planets that grow in all three 
planet traps. We adopt the above analytical prescriptions for planetary growth and migration. We summarize our 
technical procedures here. The standard treatment of mass accretion and planetary growth is given in 
Appendix \ref{app2}.

When the mass of protoplanets is less than $M_{mig,I}$, their mass increases with time following the standard 
oligarchic growth (see equation (\ref{growth1})) while their semi-major axes remain roughly the same. Time evolution 
also reduces the surface density of disks (gas and dust), and hence the growth rate also changes with time (see equation 
(\ref{tau_cacc})). Once they acquire masses that are larger than $M_{mig,I}$, they start to migrate inward. When they 
are at their planet traps, they move inward at the same rate as their traps. If they are left behind, they quickly 
catch up with their traps due to the standard rapid type I migration, and then follow the movement of the traps. If 
their planet traps disappear due to convergence with other traps, it is assumed that the planets follow new planet 
traps that survive the convergence. If the planets become more massive than the critical mass of cores above which 
their envelopes cannot maintain hydrostatic equilibrium (see equation (\ref{m_ccrit})), then accretion of gas onto 
the cores begins. The gas accretion rates are regulated purely by the mass of cores (equation (\ref{growth2})). Through 
our experiments, we find that, for most cases, core formation is completed when they are captured in their traps.  

When a planet's mass reaches the gap opening mass, it undergoes the standard type II migration. This results in 
"dropping-out" of the trapped planet from its planet trap and happens because the planet is now too massive to open 
up a gap in the disk (which leads to different orbital evolutions between the planet and the planet trap). When the 
planets are within the 
dead zone, type II migration becomes slower through a low value of $\alpha(=\alpha_D$) while, for the planets outside 
the dead zone, the value of $\alpha_A$ is used for the migration. If the planet attains the mass of $f_{max}M_{max}$, 
where $f_{max}$ is a controllable parameter (see Appendix \ref{app2}), its accretion is terminated. $M_{max}$ is a 
decreasing function of time (see equation (\ref{M_max})), so that planets that need a long time to grow up to gas 
giants tend to be less massive while planets that can quickly become gas giants tend to be more massive. Even when 
planet formation is largely complete, their disks may still have a sufficient amount of gas to drive type II migration. 
In this case, the type II migration is slowed down by the inertia of the planets. The accretion rate $\dot{M}$ declines 
with time, and at certain time $\dot{M}$ becomes equal to photoevaporation rates $\dot{M}_{pe}$. When this is satisfied, 
the positions of the planets freeze in the mass-semi-major axis diagram.  

\section{Results} \label{resu}

We are now in the position to discuss the results of evolutionary tracks of planets that grow in disk inhomogeneities. 
As shown in Fig. \ref{fig4}, most evolutionary tracks behave similarly. Therefore, we first discuss the results of 
the dead zone in detail (see $\S$ \ref{resu1}), and then examine all the three disk inhomogeneities (see $\S$ 
\ref{resu2}). We compare the results with the observations in $\S$ \ref{resu3}.   

\subsection{Planetary growth in a dead zone trap} \label{resu1}

\begin{figure}%[!ht]
\begin{center}
\includegraphics[width=9cm]{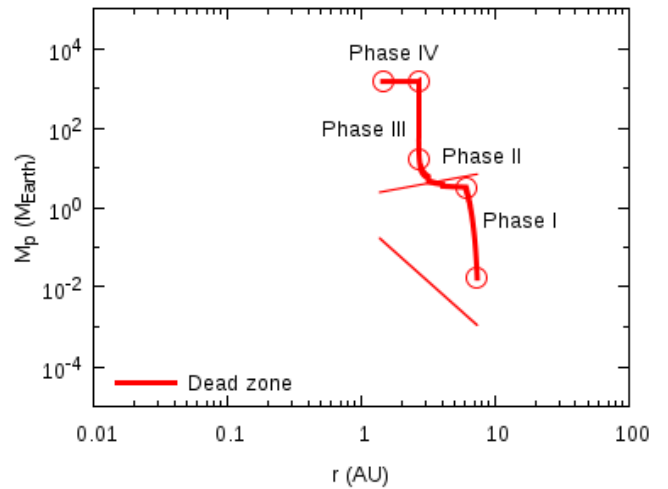}
\caption{An evolutionary track of a planet that grows in a dead zone trap. The track (denoted by the thick line) can be 
divided into four phases. In Phase I, core formation takes place very rapidly in $\tau \sim 10^{5}-10^{6}$ years, which 
is much faster than the radial movement of the trap at that time. This results in largely vertical motion in the 
diagram. In Phase II, the core accretes gas onto its envelope. Its timescale is very slow ($\sim 2\times 10^{6}$ years 
in this case). Therefore, it moves horizontally in this diagram. The mass of the core in Phase I and most of Phase II 
is within the trapping regime that is represented by the upper and lower thin lines. Toward the end of Phase II, the 
core drops-out from the trap by opening up a gap in the disk and undergoing type II migration. Phase III is runaway gas 
accretion onto the core. The timescale of this phase is very short ($< 10^{5}$ years). As a result, it moves vertically 
in this diagram. Planet formation completes during Phase I to III. In Phase IV ($\gtrsim 10^{6}$ years), the gas giant 
moves inward due to type II migration that is slowed down by the inertia of the planet. When photoevaporation of the 
gas disk becomes important, type II migration is terminated and its final radial position and orbital period are 
obtained.}
\label{fig3}
\end{center}
\end{figure}

The evolutionary track of a growing planet consists of four distinct phases in the 
mass-semi-major axis diagram (see Fig. \ref{fig3}). The first phase is formation of cores of gas giants through runaway 
and oligarchic growth \citep[e.g.][]{ws89,ki98}. The mass of the core in this phase is high enough to keep up with the 
movement of the trap while the torque of the core acting on the disk is too weak to open up a gap there. Thus, the core 
remains within a trapping regime and follows its movement. The timescale of this phase is order of $\sim 10^{5}-10^{6}$ 
years, which is much shorter than the disk lifetime ($\tau_{disk} \sim 8.8 \times 10^{6}$ years in this setup) as shown 
in previous studies \citep{ki02}. Hence, the protoplanet moves upwards in mass while moving little in orbital radius or 
period. 

As the feeding zone empties, core formation is terminated and the second phase begins, wherein gas accretion 
onto the envelope occurs. It was well known that the timescale of this phase was problematically long for earlier models 
\citep[$\gtrsim 10^{7}$ years,][]{p96}. However, recent studies improved the previous models and revealed that the 
timescale is highly sensitive to the optical depth of the envelope. For these realistic conditions, it is significantly 
shorter than the disk lifetime \citep{lhdb09}. In our calculation, this timescale is about $2\times10^{6}$ years and 
hence the core still has sufficient time to finally grow up to a gas giant within $\tau_{disk}$. The core during most 
of this phase is trapped. As a result, its radial evolution is mainly determined by the slow movement of the dead zone 
trap, and the protoplanet moves to shorter radii and periods while at nearly a constant mass. Toward the end of this 
phase, the core becomes massive enough, so that the tidal torque it exerts upon the disk becomes comparable to the 
viscous torque that evolves gas disks, leading to gap formation in the disks and type II migration of the planetary 
core.

When the mass of the gaseous envelope cannot be supported by the gas pressure, runaway gas accretion onto the core 
takes place (Phase III). The timescale of this phase is very short ($\lesssim 10^{5}$ years), and consequently its 
evolutionary path is almost vertical in the mass-semi-major axis diagram. These three successive phases are the main 
path to forming gas giants in the core accretion scenario \citep{p96,lhdb09}. The massive planet opens up a gap in 
the disk and undergoes type II migration. This switch from type I to type II migration results in "dropping-out" of 
the planet from the moving trap and decouples it from the movement of the planet trap. 

The onset of Phase IV completes 
the formation of a gas giant. During this phase ($\gtrsim 10^{6}$ years), type II migration moves the gas giant inward 
further. However, this process is minimized by the inertia of the massive planet \citep{sc95,ipp99}. 

Planets arrive at their final position in the mass-semi-major axis diagram when the disk is finally dissipated. 
Photoevaporation of the disk by high energy radiation from the central star is likely to be the dominant mechanism of gas 
dispersal in the disks \citep[e.g.][]{gh09}, and will terminate type II migration. As a result, the final orbital period 
and mass of the planet are achieved. Thus, Fig. \ref{fig3} summarizes how concurrent evolution of planetary growth and 
migration proceeds in the mass-semi-major axis diagram: a core is formed in a dead zone trap that is initially located 
at $\sim$ 7 AU. Following the movement of the dead zone trap, the core is transported to $\sim$ 3 AU. Simultaneously, 
it undergoes the two main phases of gas giant formation. The completion of the final runaway gas accretion onto the 
core and subsequent type II migration involve further evolution of the planet in the diagram. When photoevaporation 
becomes important, the gas disk is removed and the position of the planet in the diagram "freezes-out".  

\subsection{Planetary growth in all the three planet traps} \label{resu2}

\begin{figure}%[!ht]
\begin{center}
\includegraphics[width=9cm]{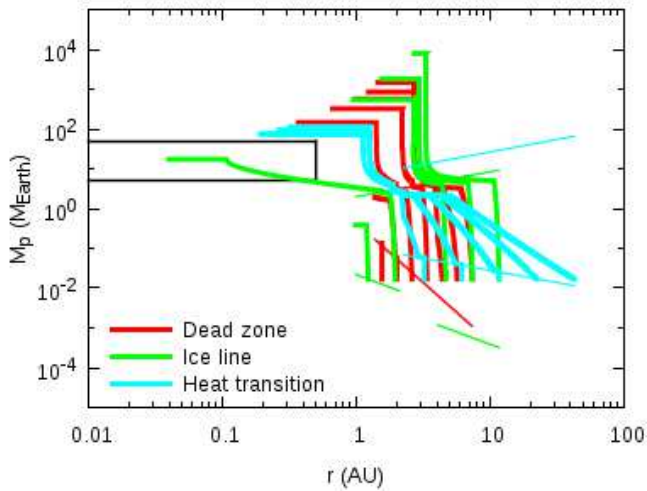}
\caption{Evolutionary tracks of planets that grow in all three planet traps. The tracks for the dead zone are denoted 
by the red thick lines, the ice line by the green, and the heat transition by the light-blue. Corresponding thin lines 
represent the trapping regimes. Different tracks correspond to planetary growth that initiates at different times (see 
Table \ref{table5}). The transport mechanism of cores by planet traps plays the crucial role in producing the 
mass-period relation; low mass cores that need longer time to grow are more likely to be transported toward smaller 
orbital radii while massive cores that can readily drop out of the moving traps tend to distribute further away from 
the star. Thus, there are distinct populations that arise from the difference in the properties of the planet traps and 
the resultant planetary growth, which results in the trend that planetary mass increases with period. Earlier papers, 
\cite{il04i,il08v} predicted a planet desert demarcated by the black rectangle. We emphasize that our model predicts 
the presence of planets there. }
\label{fig4}
\end{center}
\end{figure}

Fig. \ref{fig4} shows the computed evolutionary tracks of planets that grow at all three disk inhomogeneities. 
Different 
lines at each planet trap correspond to different evolutionary tracks in which planetary growth starts at different 
times (see Table \ref{table5}). Despite the difference in the starting time (and position), most planets formed at the 
dead zone and heat transition traps end up at $r \sim 1$ AU ($\sim 500$ days) and $r\sim 0.1$ AU ($\sim 10$ days), 
respectively. At the heat transition trap, the surface density of dust is low. Therefore, cores that grow there spend 
a long time in the trapping phases (Phase I and II). This maximizes the distance over which cores are transported and 
results in the distribution of cores that hover preferentially around $\gtrsim 1$ AU. Since the low mass cores get 
distributed over smaller orbital radii and less time remains for the cores to grow up to gas giants, they finally 
remain less massive ($\lesssim 100 M_{\oplus}$), and are located around smaller orbital radii ($\sim 0.1$ AU). The 
same argument is applied to planets formed in the dead zone trap. However, the surface density of dust at the dead zone 
is considerably higher than that at the heat transition. Consequently, the final mass of cores trapped at dead zones 
becomes larger, core formation completes earlier, and the distribution of cores is shifted to $\sim 3$ AU. These 
combined differences result in the populations of more massive planets orbiting at $\sim 1$ AU. 

The evolutionary tracks associated with protoplanets carried by the ice line trap show some differences. The resultant 
planetary population spreads out over a wider range in the mass-semi-major axis diagram (see Fig. \ref{fig4}). 
Nonetheless, this can be also understood by the the surface density of dust and the resultant core formation there. At 
the ice line, the surface densities are substantially higher than that at the dead zone and heat transition and hence 
the formation of cores is most efficient. This typically results in most massive cores. At the early stage of disk 
evolution, therefore, the most massive cores are preferentially formed there. They can readily drop-out from the moving 
trap and pile up around larger orbital radii ($r \sim 5$ AU). These massive cores at larger orbital radii lead to 
the formation of more massive gas giants that finally orbit at $\gtrsim 1$ AU. In the later stage of disk evolution, 
the high dust densities at the ice line can still form cores while at that time the other traps not due to lower dust 
density there. This is the physical reason of the wide spread of planetary population due to the ice line traps.  

\subsection{Comparisons with the observations} \label{resu3}

\begin{figure}%[!ht]
\begin{center}
\includegraphics[width=9cm]{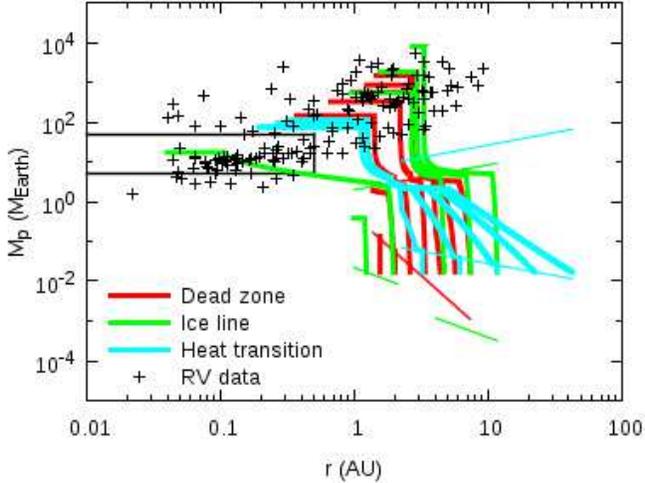}
\caption{Comparisons with the observations. The observational data are adopted from \citet{mml11} (as Fig. \ref{fig1}). 
Our theoretical mass-period relation is consistent with the observations. Also, the presence of many observed low mass 
planets ($\lesssim 50 M_{\oplus}$) at $r\lesssim0.5$ AU provides further support on our model.}
\label{fig5}
\end{center}
\end{figure}

We now compare our results with the observations. As already presented in Fig. \ref{fig4}, our model shows that 
the superposition of all tracks for planets that grow in three planet traps constitutes a theoretical mass-period 
relation, wherein the final distribution of the mass of the planets is an increasing function of their periods. 
This is consistent with the observed mass-period relation, as the observational data scatter around the locus 
of end points of our tracks (see Fig. \ref{fig5}). 

This is one of the most important findings in this paper. 
As discussed in $\S$ \ref{resu2}, this arises from the fact that there are considerable differences in the properties 
of the planet traps that regulate planet formation and migration. As a result, different planet traps have different 
preferred loci at which evolutionary tracks end up in the mass-semi-major axis diagram. Thus, planet traps 
act as a filter for distributing cores - massive cores readily drop out from moving traps and tend to orbit further 
away from the central star while low-mass cores are trapped for a long time and tend to orbit close to the host star - 
and play the central role in generating the theoretical mass-period relation.

In addition, the prediction that distinct sub-populations can arise depending on the trapping mechanism has several 
observational consequences. For example, our model provides a physical explanation for the observed pile up of gas 
giants at $\sim 1$ AU. This again relies on the argument that planet formation efficiency highly depends on the surface 
density of dust at planet traps. At the dead zone and ice lines, the dust density is expected to be high due to the 
low disk turbulence, and hence planet formation rates are high there. On the other hand, the formation rate would be 
low at the heat transition trap due to low dust density. This results in a general trend that more planets are readily 
formed at the dead zone and ice line traps that end up at $r \sim 1$ AU (see Fig. \ref{fig5}). 

Furthermore, our model predicts the population of low mass planets ($\lesssim 50 M_{\oplus}$) with $r\lesssim 0.5$ AU. 
This arises from planet formation that takes place in the moving ice line trap (see Fig. \ref{fig5}). Even in the later 
stage of disk evolution, the highest dust density there enables the formation of low-mass planets that end up in the 
desert. On the contrary, the most advanced population synthesis models predict a planet desert there 
\citep[also see the footnote 3 in $\S$ \ref{intro}]{il04i,il08v}. The presence of the many observed exoplanets in the 
region agrees well with our findings.

\begin{figure}%[!ht]
\begin{center}
\includegraphics[width=9cm]{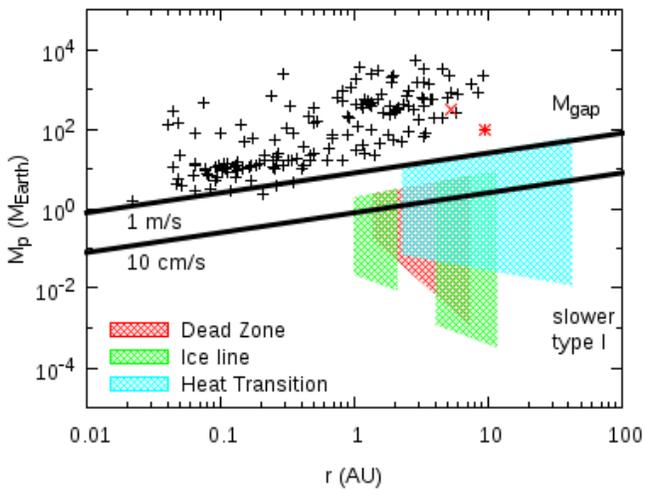}
\caption{Prediction of planet deserts for the CTTS case. The desert produced by the dead zone trap is denoted by the 
red hatched region, the ice line trap by the green, and the heat transition by the light-blue. These regions are the 
consequence of trapping of type I migrators and defined by the gap opening mass $M_{gap}$ (see equation (\ref{M_gap})) 
and the mass of planets above which they can keep up with the movement of their planet traps (see 
equation (\ref{M_typeI}), also see Table \ref{table3}).}
\label{fig6}
\end{center}
\end{figure}

Finally, our models predict the existence of planet deserts that are quite different in the mass-period space than those 
claimed by \citet{il04i,il08v}. Fig. \ref{fig6} shows our deserts, denoted by hatched regions. They are produced due to 
trapping and subsequent transport of cores. This leads to the evacuation of the cores from these regions in which 
they have initially grown up. As a result, these regions are regarded as void of planets. More specifically, we 
define our deserts by estimating the mass ranges of planets that can be captured at the planet traps and following their 
movement: $M_p < M_{gap}$ and $\tau_{mig,I} < \tau_{vis}$ (see $\S$ \ref{charact_mass}). This kind of planet desert is 
active only for gas disks. There are a number of possibilities to fill out our deserts: that successive formation 
of rocky planets after gas disks disperse may ultimately fill out the regime; that, even in the epoch of gas disks, 
planetary cores formed far beyond our deserts may eventually distribute there due to planetary migration; and that 
planet-planet scatterings induced by convergence of multiple planet traps may deliver the scattered cores into our 
deserts. Nonetheless, our predictions are valuable in a sense that such regions are the primary target of the 
current and ongoing observational surveys \citep{mml11,h11}. 

\section{Parameter studies} \label{ParaStu}

We perform parameter studies by varying disk and stellar parameters in order to examine how robust 
our findings discussed in $\S$ \ref{resu3} are. Also, we investigate the effects of the inertia of planets by changing 
the value of $f_{mig,slowerII}$ in $\S$ \ref{ParaStu3} to differentiate them from the role of planet traps discussed 
above.

\subsection{Disk parameters}

We first focus on parameters for dead zones. We have adopted the parameterized treatment for the structures of dead 
zones, wherein they are represented by $\Sigma_{A0}$ and $s_A$ (see equation (\ref{Sigma_A})). Even in the most recent 
studies, it is still somewhat uncertain what the precise structure of the dead zones is \citep{mp06,mll12}. Therefore, 
we utilize our parameter study in order to discuss how sensitive our findings are to the structures of the dead zones.

\begin{figure*}
\begin{minipage}{17cm}
%\begin{figure}%[!ht]
\begin{center}
\includegraphics[width=5.5cm]{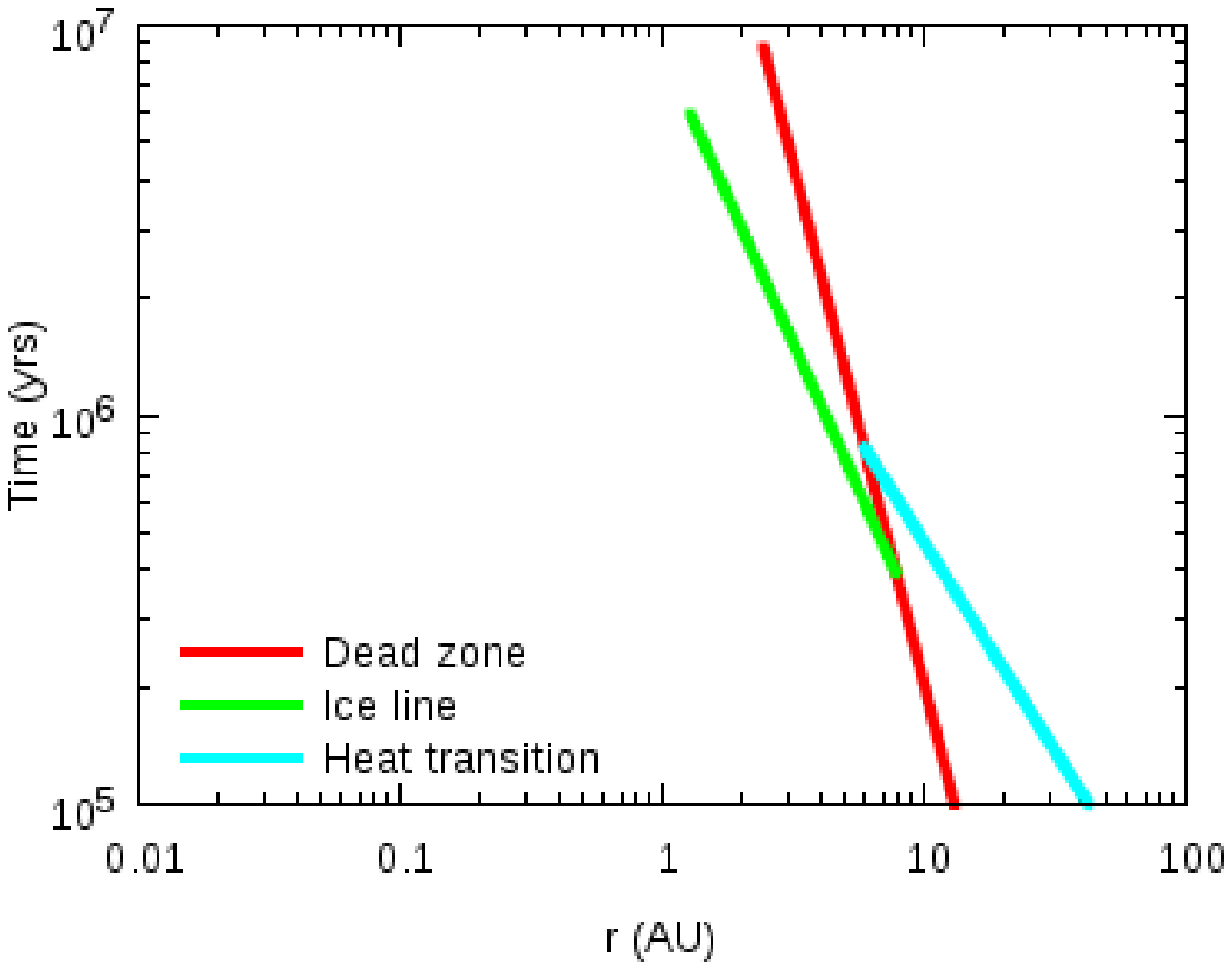}
\includegraphics[width=5.5cm]{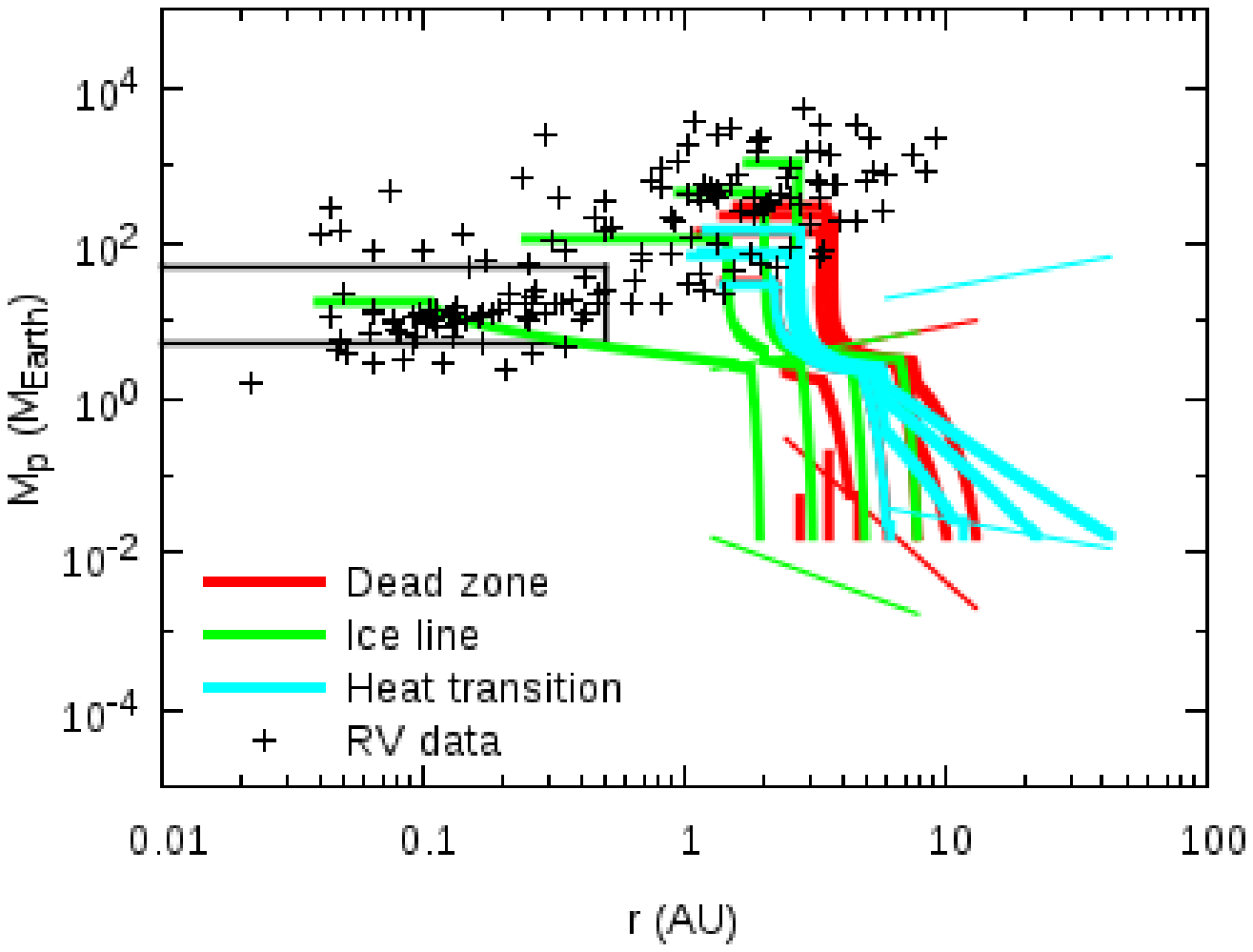}
\includegraphics[width=5.5cm]{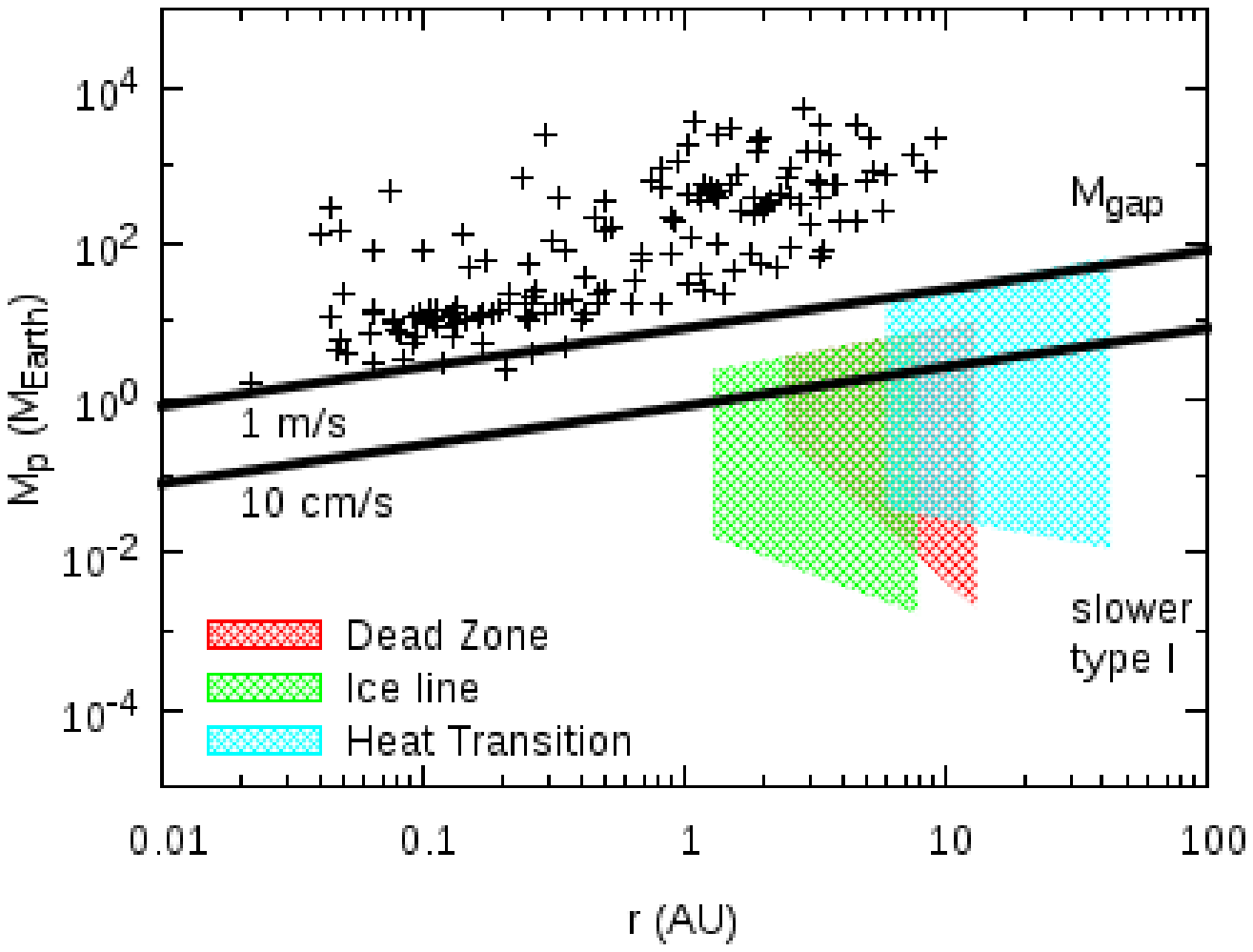}
\includegraphics[width=5.5cm]{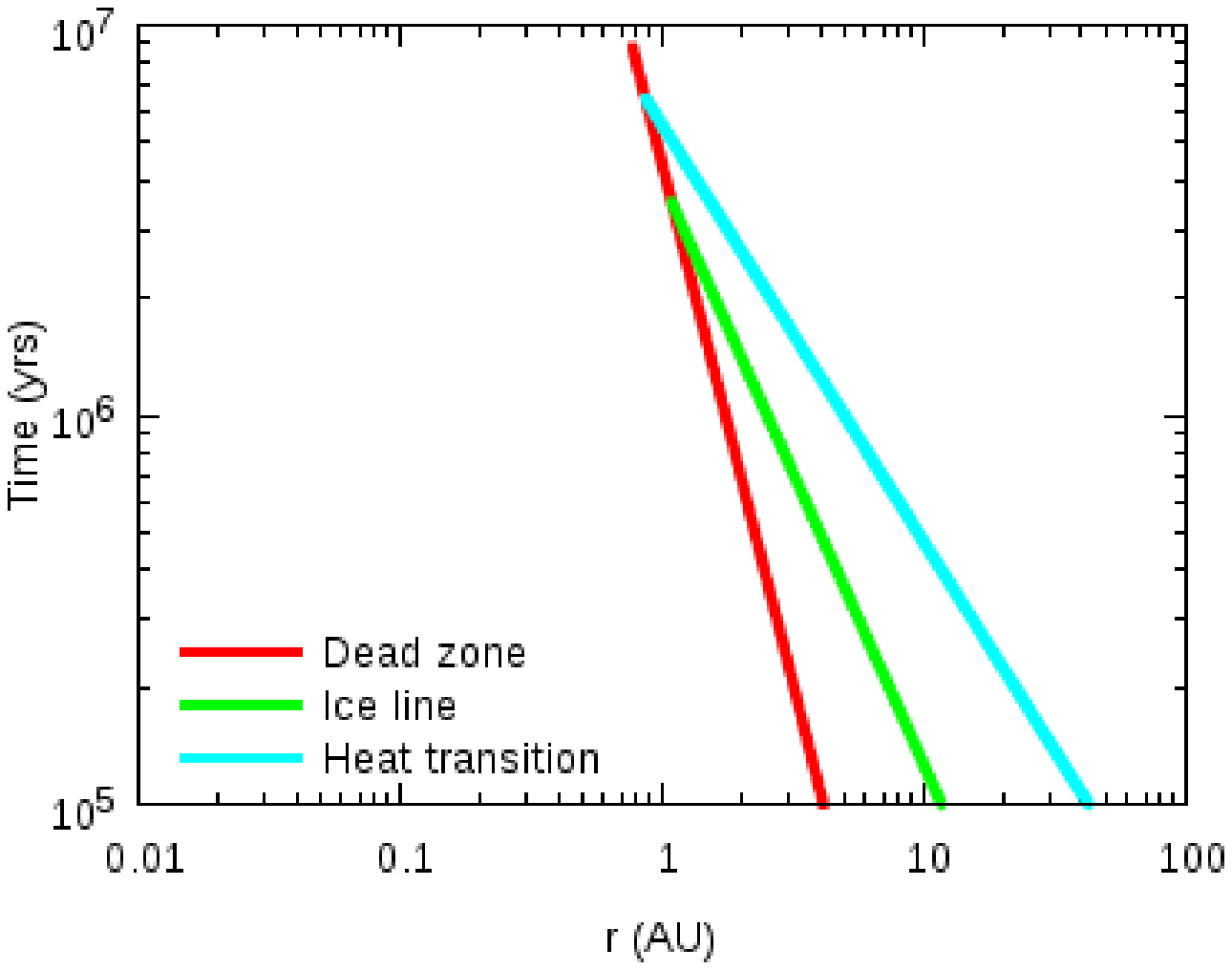}
\includegraphics[width=5.5cm]{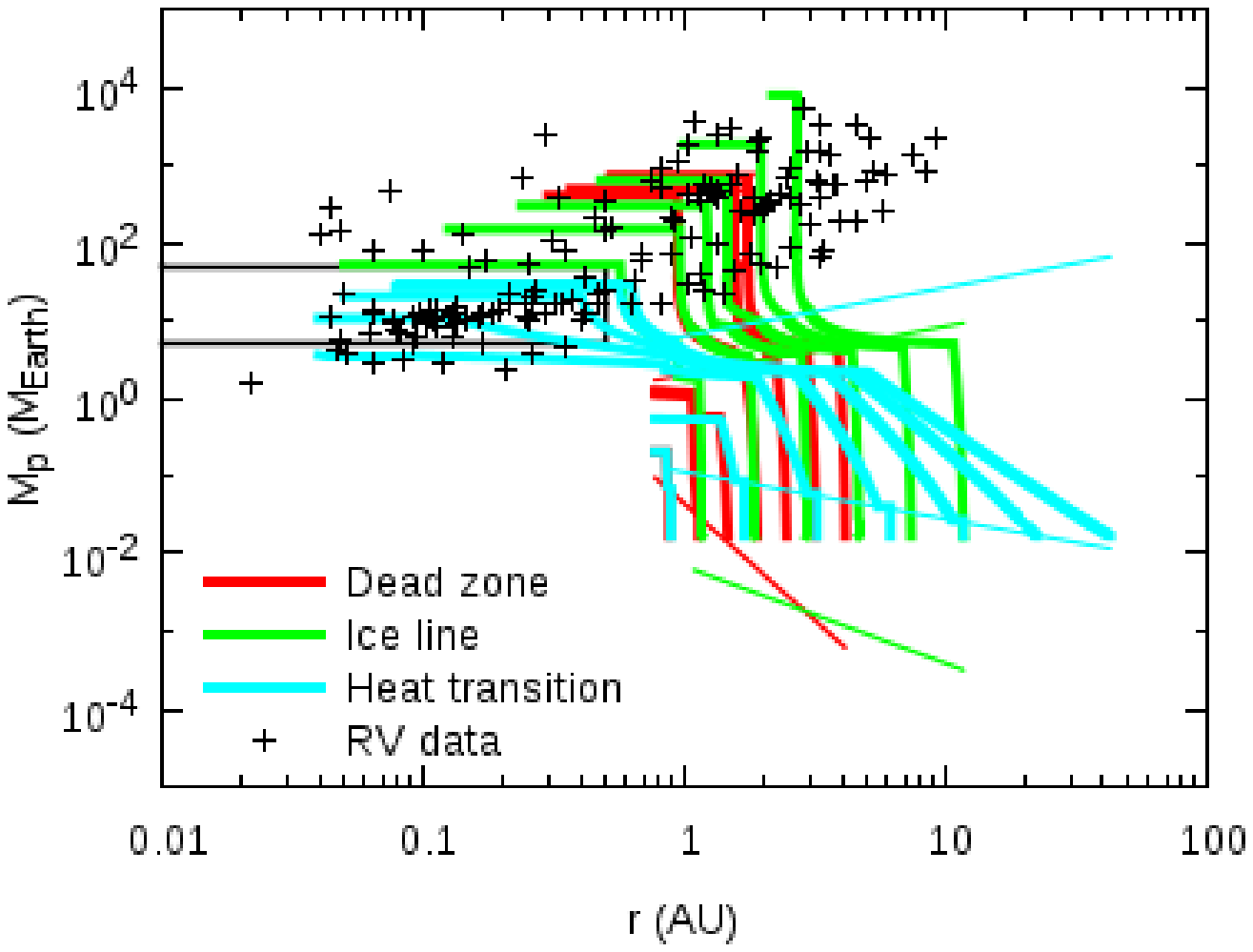}
\includegraphics[width=5.5cm]{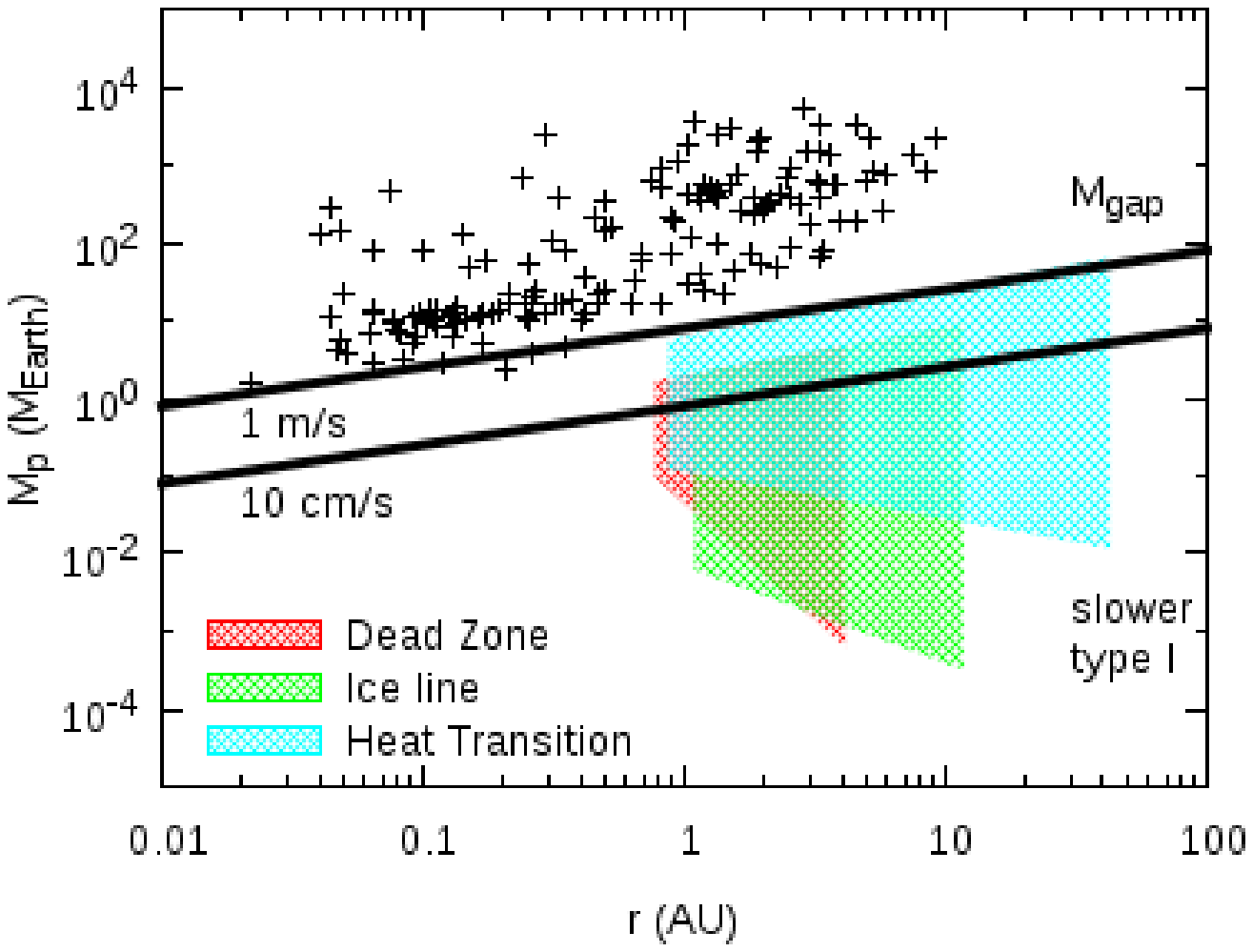}
\includegraphics[width=5.5cm]{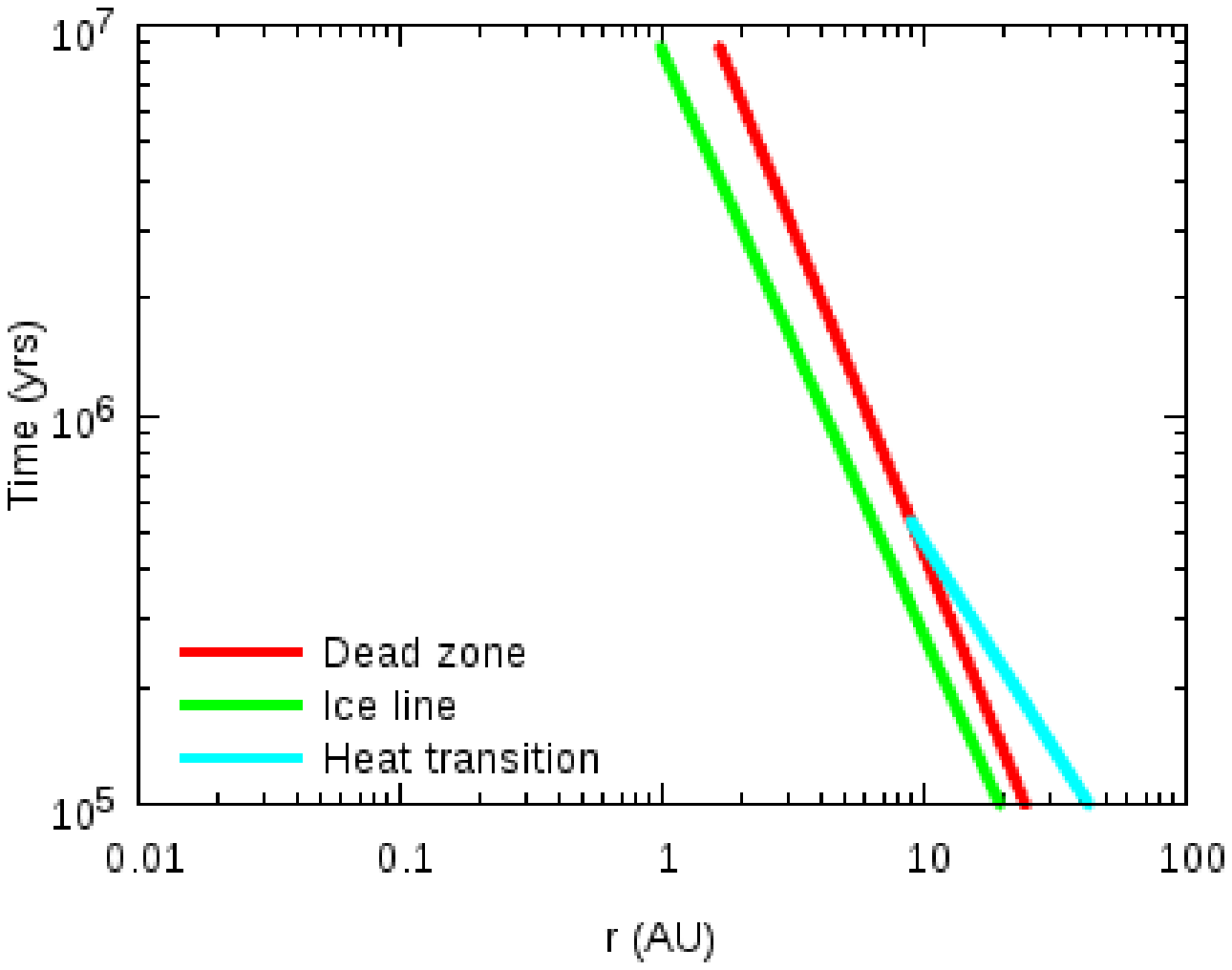}
\includegraphics[width=5.5cm]{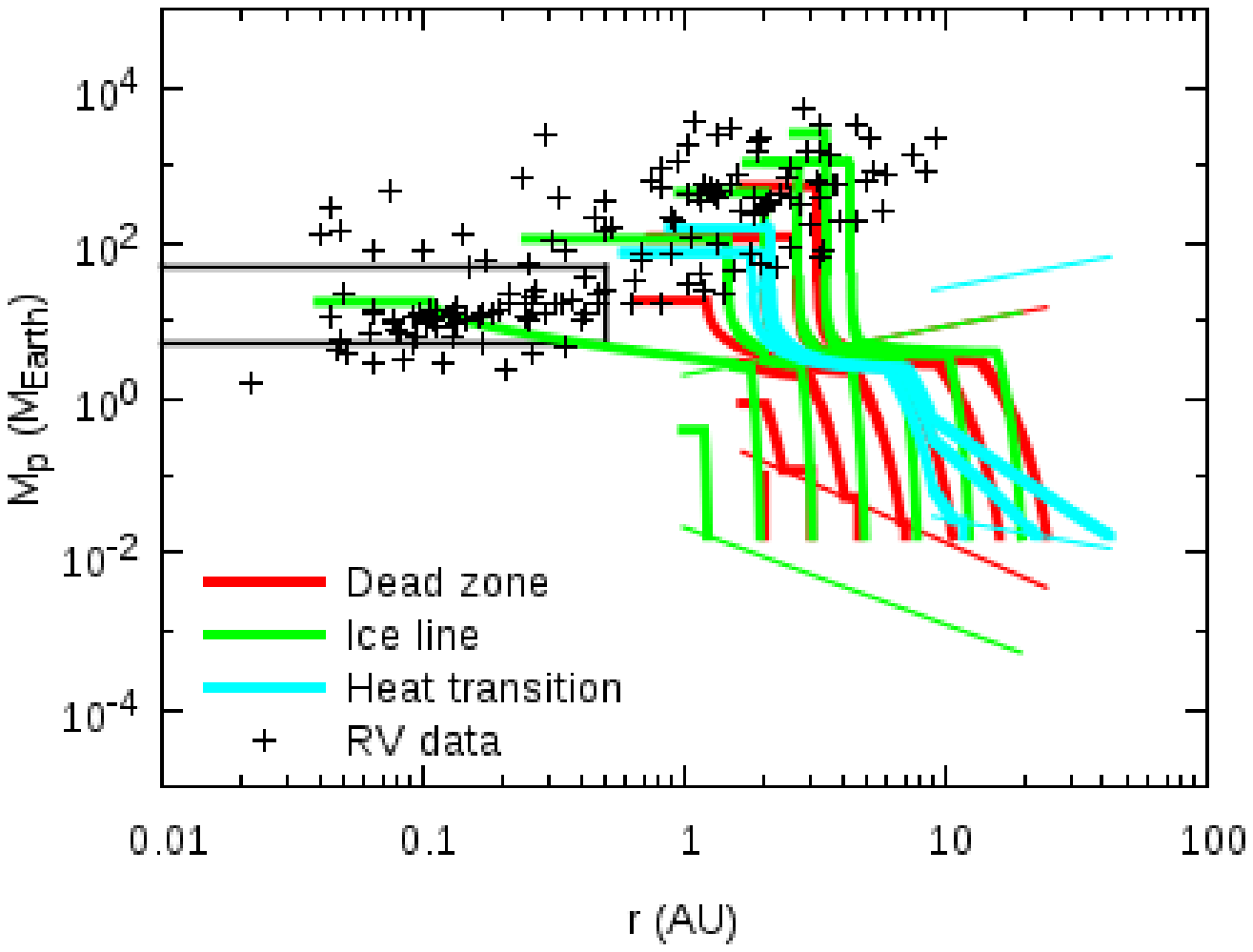}
\includegraphics[width=5.5cm]{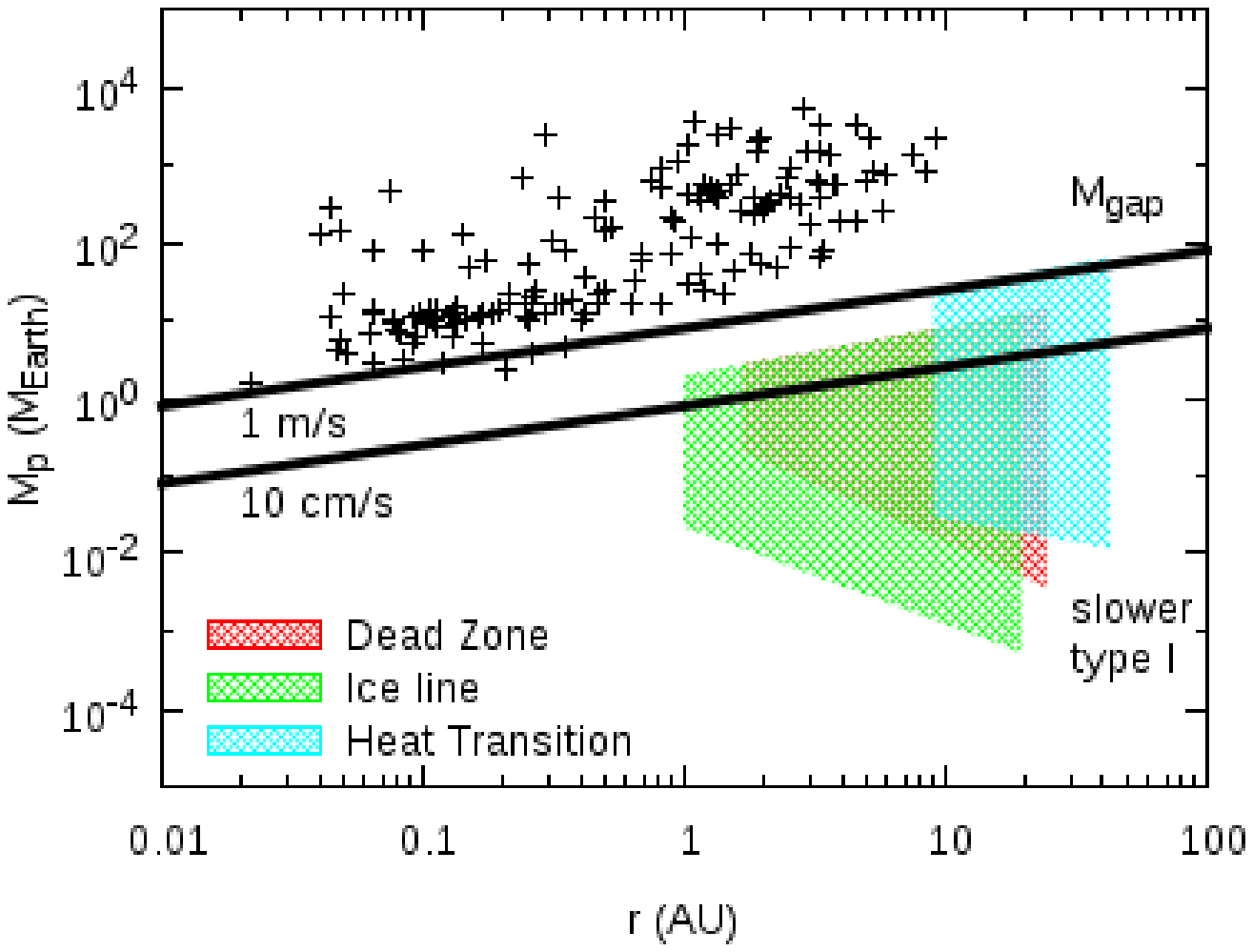}
\includegraphics[width=5.5cm]{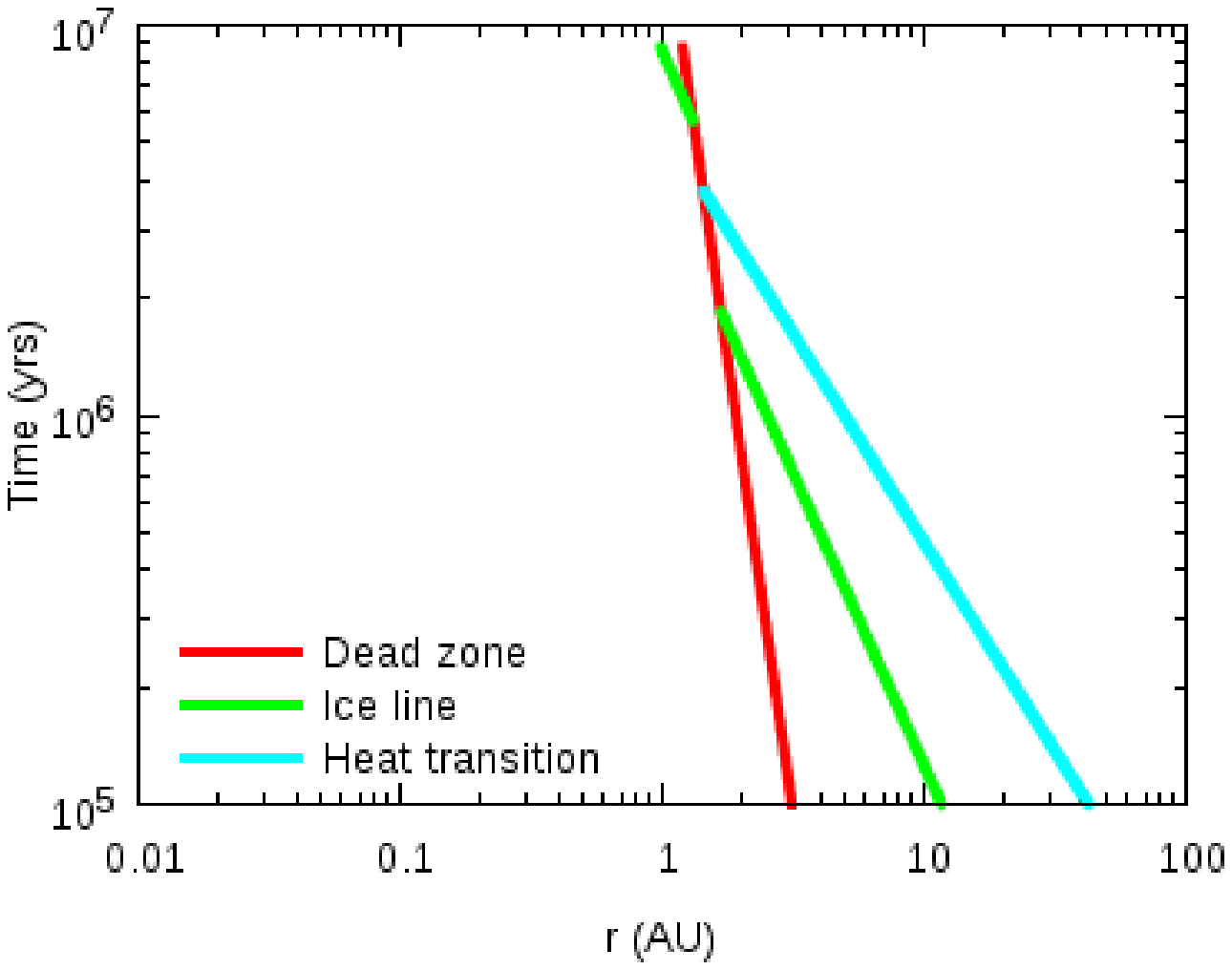}
\includegraphics[width=5.5cm]{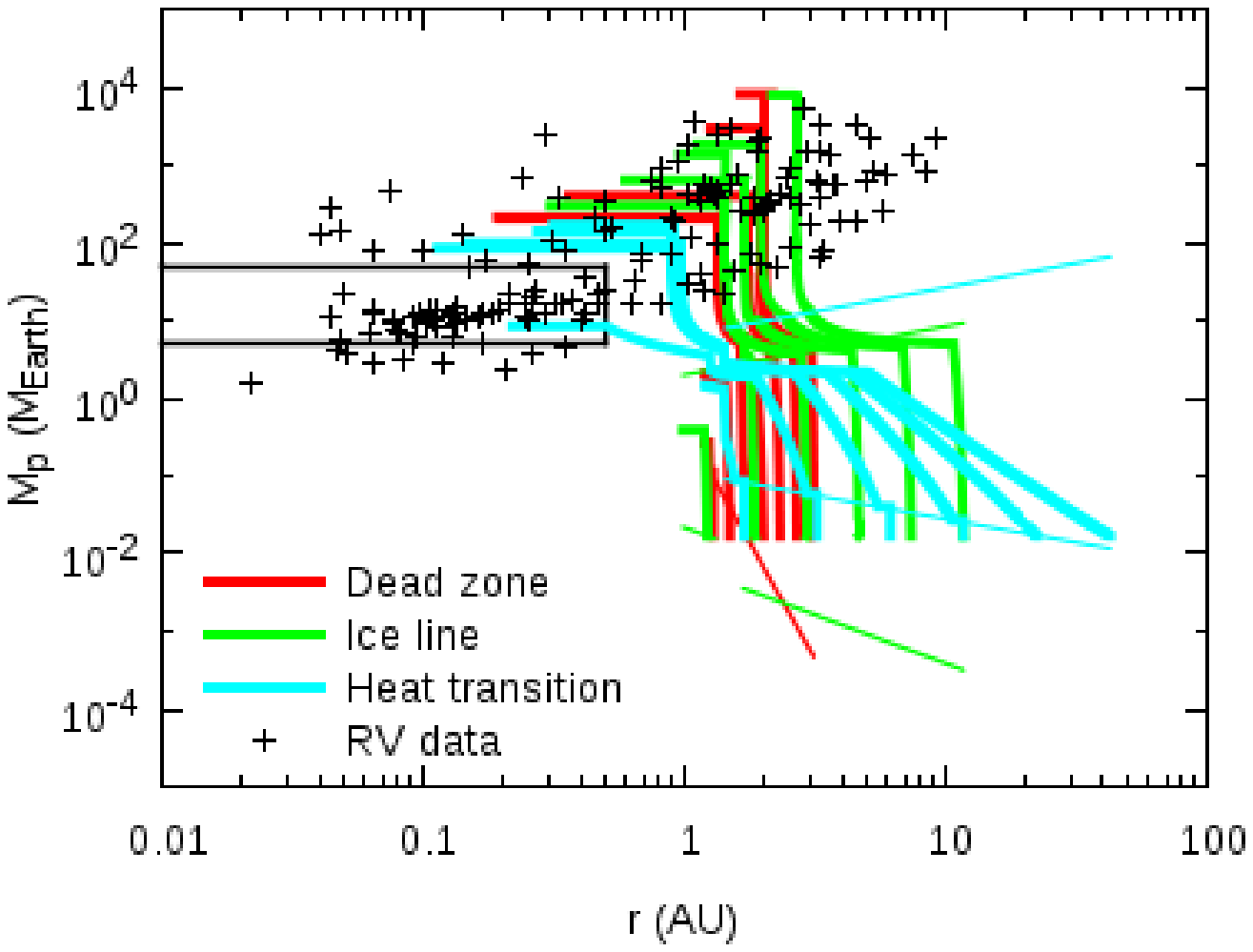}
\includegraphics[width=5.5cm]{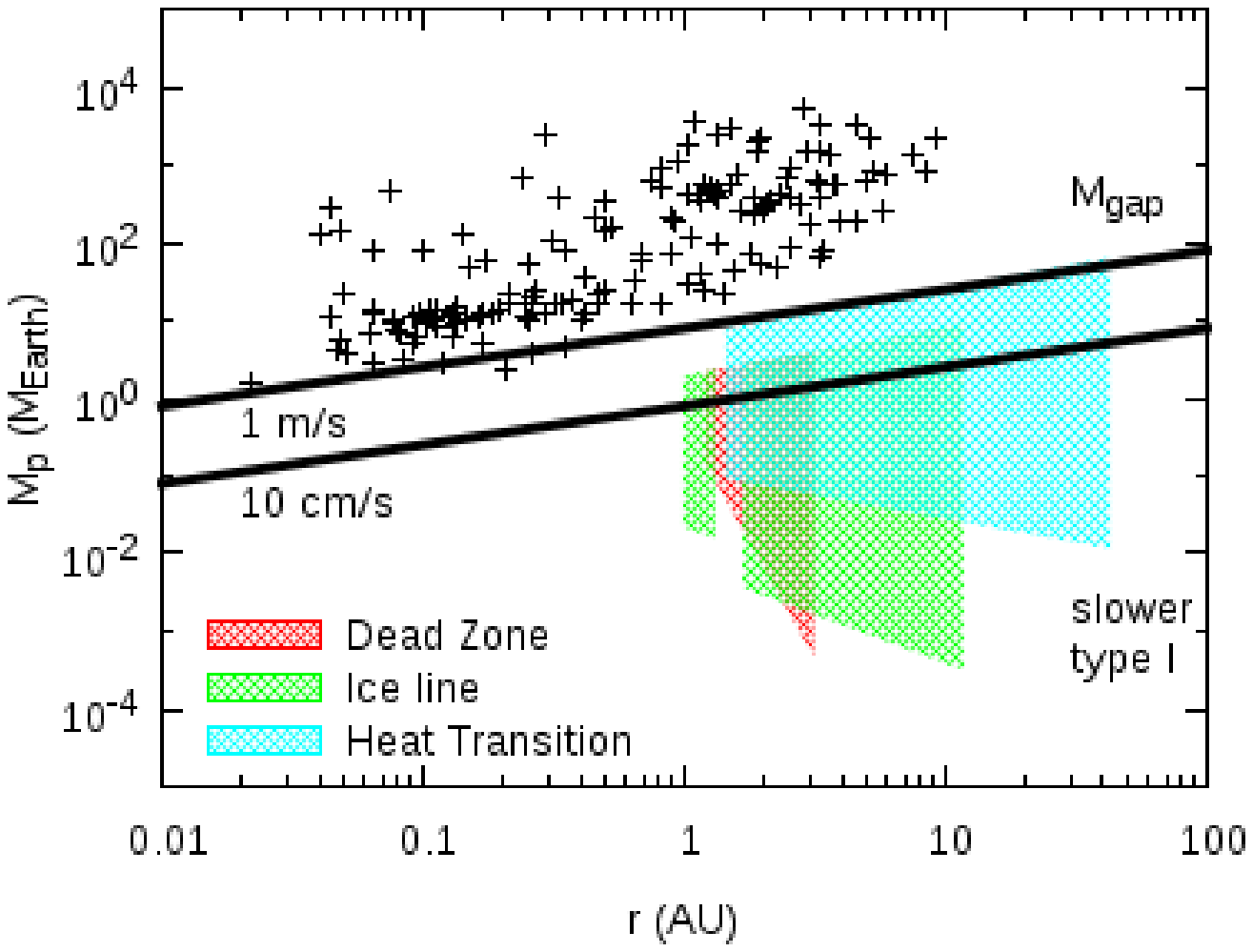}
\caption{Parameter study of dead zones (see Table \ref{table6}). The movements of three planet traps are shown in 
the left column (as Fig. \ref{fig2}), the evolutionary tracks of planets that grow there in the central (as Fig. 
\ref{fig5}), and the behaviors of the trapping regimes in the right (as Fig. \ref{fig6}). The top panels show 
the results of Run A1, the second for Run A2, the third for Run A3, and the bottom for Run A4. The results are quite 
similar 
to those of the fiducial model, and hence our important findings such as origins of the observed mass-period relation 
and the pile up at 1 AU and a prediction of low-mass planets with tight orbits are maintained for dead zones that can 
have a variety of structures.}
\label{fig7}
\end{center}
%\end{figure}
\end{minipage}
\end{figure*}

Table \ref{table6} summarizes parameters we varied. For Runs A1 and A2, the value of $\Sigma_{A0}$ is changed while 
$s_A$ varies for Runs A3 and A4. Any other parameters remain the same as the fiducial ones for all the four runs. Fig. 
\ref{fig7} shows the results of the evolution of the positions of three disk inhomogeneities (the left column), the 
evolutionary tracks of planets (the central column), and the trapping regimes (the right column). The top panels are 
for the case of Run A1, the second for the Run A2, the third for Run 3, and the bottom for Run4. One immediately 
observes 
that the results for all the four cases, especially the behaviors of the evolutionary tracks, are surprisingly similar 
to those of the fiducial case and give all three key results. Thus, we can conclude that our findings discussed in 
$\S$ \ref{resu3}, are robust even if the structures of dead zones somewhat change due to their surrounding environments 
and disk configurations.
   
\begin{table}
\begin{center}
\caption{Parameter study of dead zones}
\label{table6}
\begin{tabular}{ccc}
\hline
        &  $\Sigma_{A0}$ (g cm$^{-2}$)  &  $s_A$      \\ \hline
Run A1  &  2                            &  3          \\
Run A2  &  200                          &  3          \\
Run A3  &  20                           &  1.5        \\
Run A4  &  20                           &  6          \\
\hline
\end{tabular}
\end{center}
\end{table}

\subsection{Effects of Stellar mass}

The above parameter study leads to the conclusion that disk inhomogeneities and the resultant multiple planet traps 
play the crucial role in reproducing the key properties of observed exoplanets. Nonetheless, there are few populations 
of planets that are not covered by the fiducial model: gas giants orbiting at $r \gtrsim 5$ AU. We now examine whether 
or not this population is also predicted by our model. In order to proceed, we change the stellar mass from 0.5 to 
$0.9M_{\odot}$ and keep other parameters the same as the fiducial ones. The main motivation for changing the stellar 
mass is that the observational data are obtained from low- to high-mass stars that cover F, G, and K stars.  

\begin{figure*}
\begin{minipage}{17cm}
%\begin{figure}%[!ht]
\begin{center}
\includegraphics[width=5.5cm]{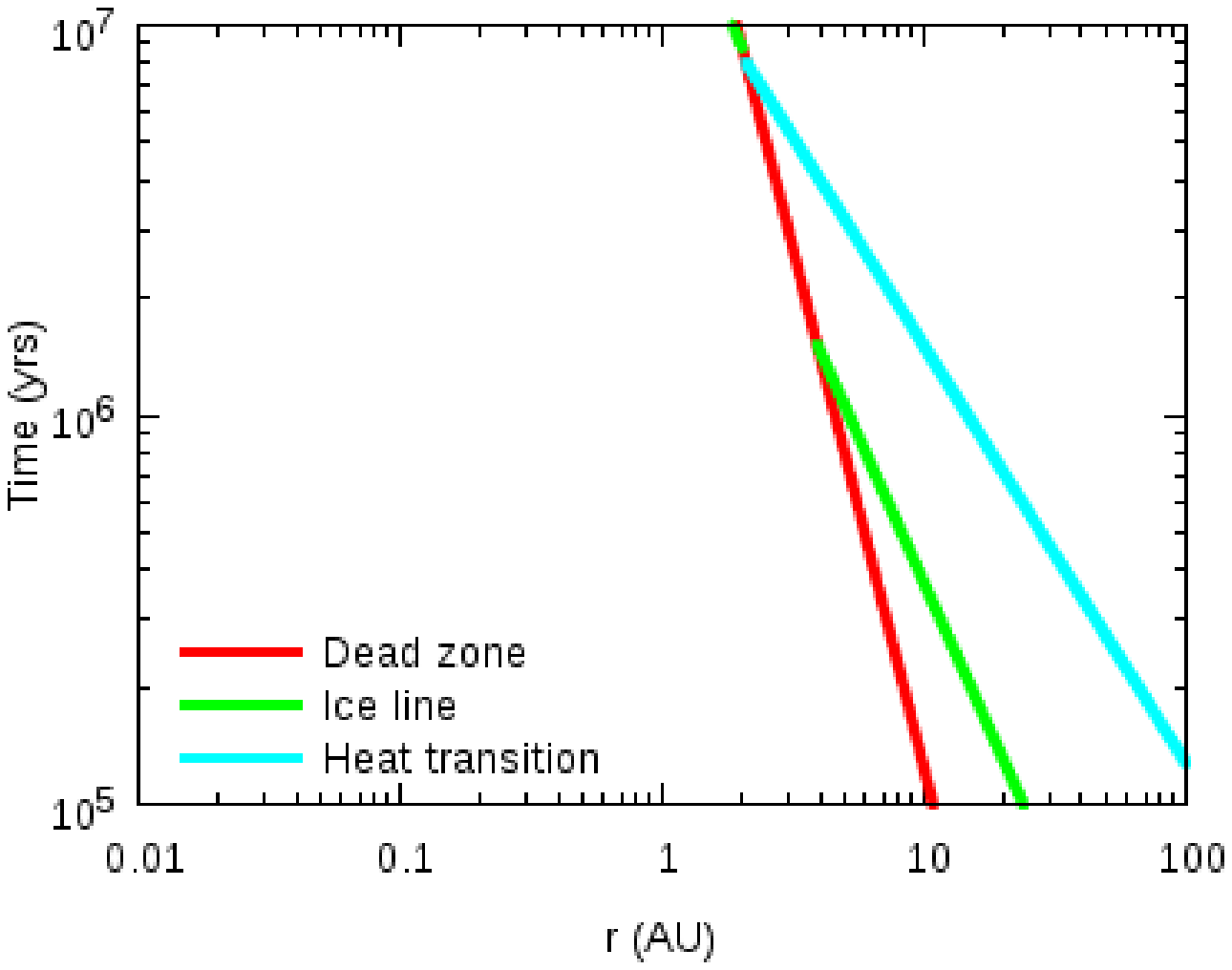}
\includegraphics[width=5.5cm]{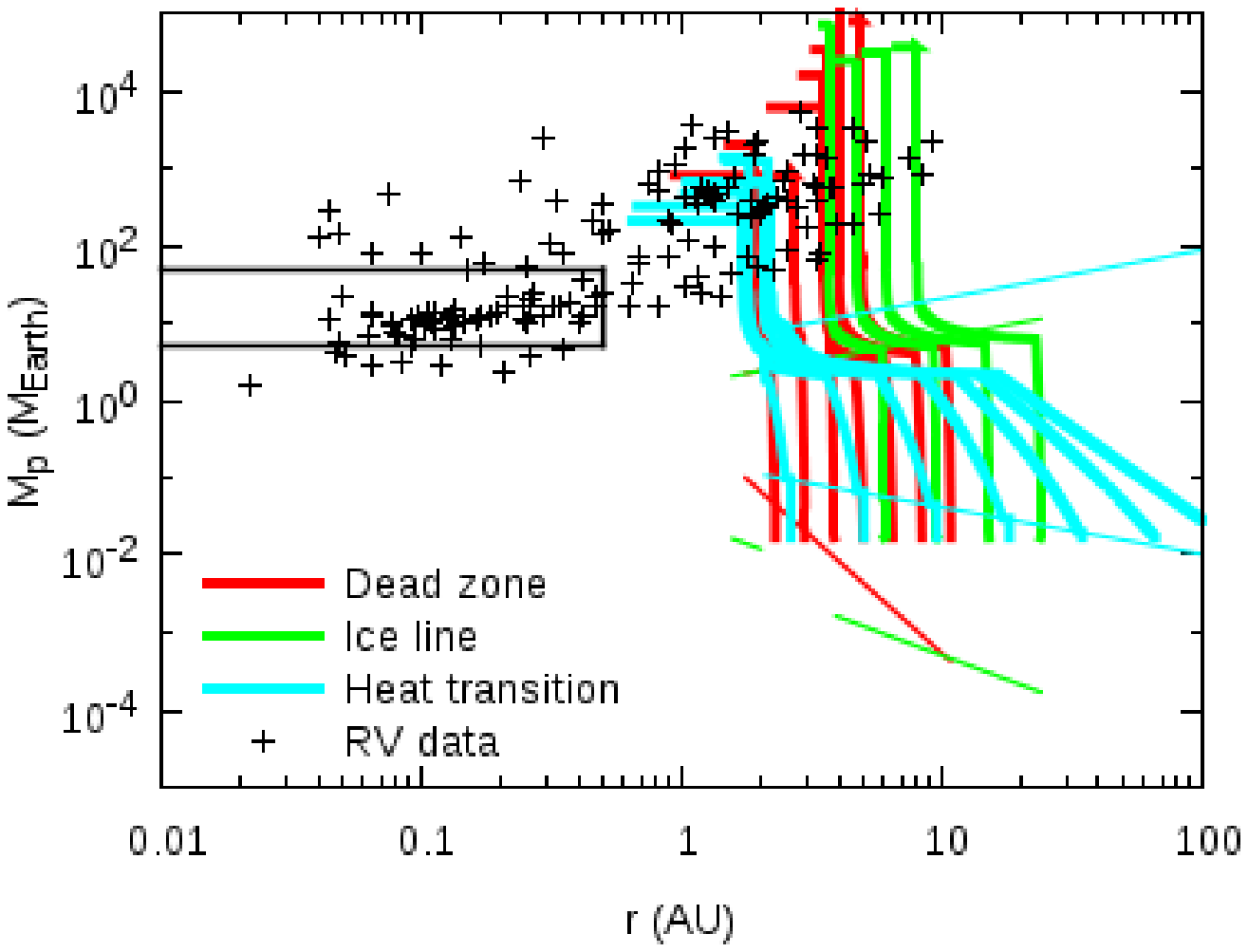}
\includegraphics[width=5.5cm]{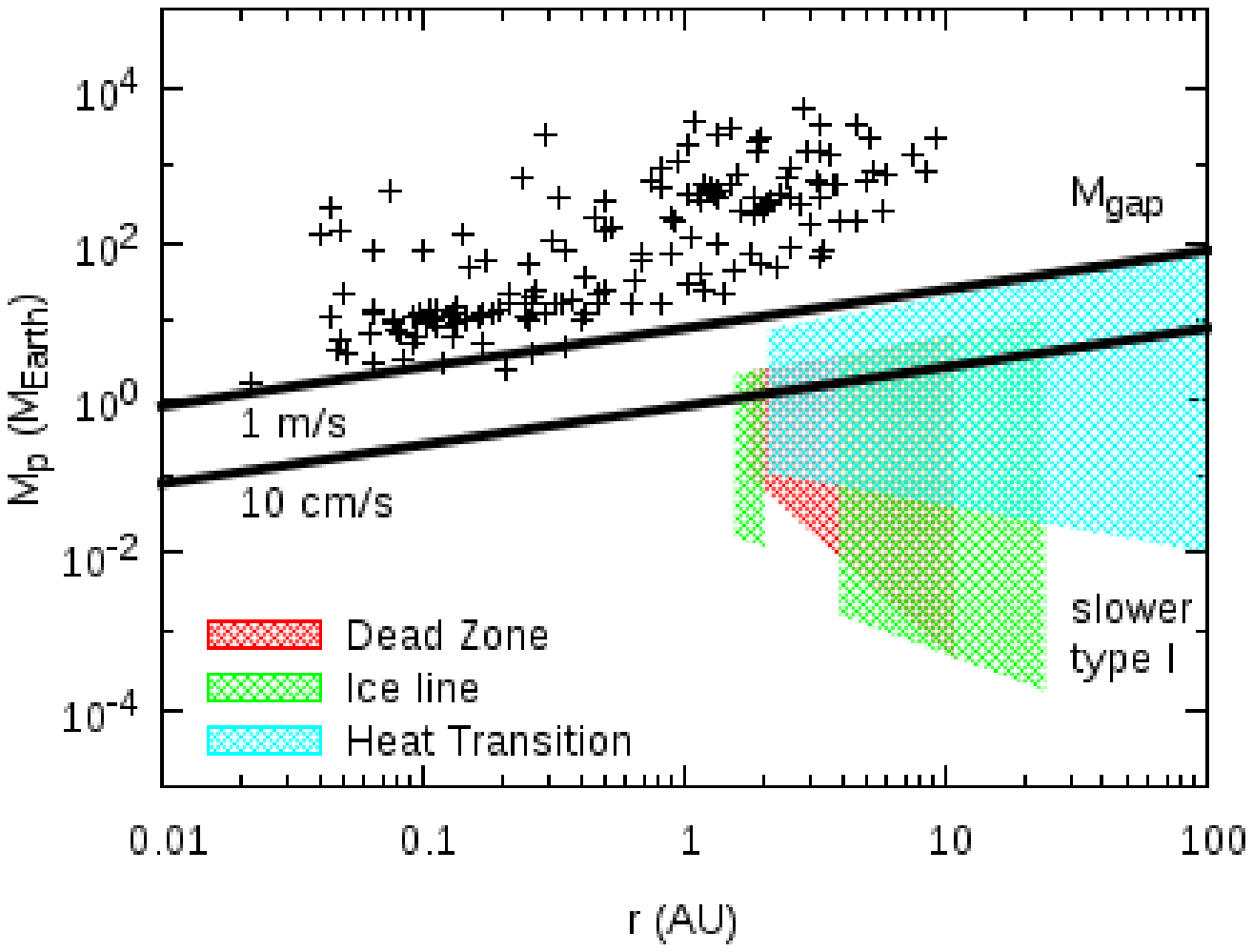}
\caption{Parameter study of the variation of stellar mass. The movements of three planet traps are shown in the left 
panel, the evolutionary tracks of planets that grow there in the central, and the behaviors of the trapping 
regimes in the right (as Fig. \ref{fig7}). The stellar mass is set as $0.9M_{\odot}$, otherwise the values of the 
fiducial model are taken. The population of gas giants ($r \gtrsim 5$ AU) not covered by the fiducial model, wherein the 
stellar mass is 0.5 $M_{\odot}$, is now reproduced. This suggests that the full extent of the data is explained by our 
model with a modest range of stellar masses, which is part of the data.}
\label{fig8}
\end{center}
%\end{figure}
\end{minipage}
\end{figure*}

Fig. \ref{fig8} shows the results of the movement of the disk inhomogeneities, the evolutionary tracks of planets that 
are formed in the traps, and the locations of the planet deserts. This figure confirms that the planets not covered in 
our previous setup ($10^{2} M_{\oplus} \lesssim M_p \lesssim 5 \times 10^{3} M_{\oplus}$ and 5 AU 
$\lesssim r \lesssim$ 10 AU) can indeed be reproduced. This is because high-mass stars result in high accretion rates 
(see equation (\ref{mdot_vis})), which corresponds to the situation that disks have high mass. As a result, planet 
formation efficiencies at all three planet traps become high, and most formed planets readily attain high mass, which 
ends up with planets distributing further away from the central star. Thus, this finding indicates that the full 
range of the statistical properties of exoplanets can be understood by our model. 

\subsection{Role of slower type II migration} \label{ParaStu3}

Based on the above discussion, a number of the fundamental statistical properties of observed exoplanets are most 
likely to be explained by multiple planet traps that capture and transport planetary cores, depending on planetary mass. 
As shown in Fig. \ref{fig4}, however, slower type II migration also drives radial drifts in the mass-semi-major axis 
diagram, following evolutionary tracks of growing planets.\footnote{The standard type II migration 
also plays some role in changing the radial distribution of planets. However, its effect is minimal, because runaway 
gas accretion proceeds so rapidly that, once planets obtain the gap-opening mass, they immediately achieve their mass 
above which the inertia of the planets is effective. As a result, the time interval for the standard type II migration 
is short enough to neglect the effects.} In addition, the slower type II migration is a function 
of planetary mass (see equation (\ref{slower_typeII})). Thus, they imply that a combination of planet traps and slower 
type II migration (not only planet traps) may play a central role in reproducing the observations. In order to examine 
the effects of the slower type II migration on our results, we carry out a parameter study in which the value of 
$f_{mig,slowerII}$ changes (see Table \ref{table7}).

\begin{figure*}
\begin{minipage}{17cm}
%\begin{figure}%[!ht]
\begin{center}
\includegraphics[width=8cm]{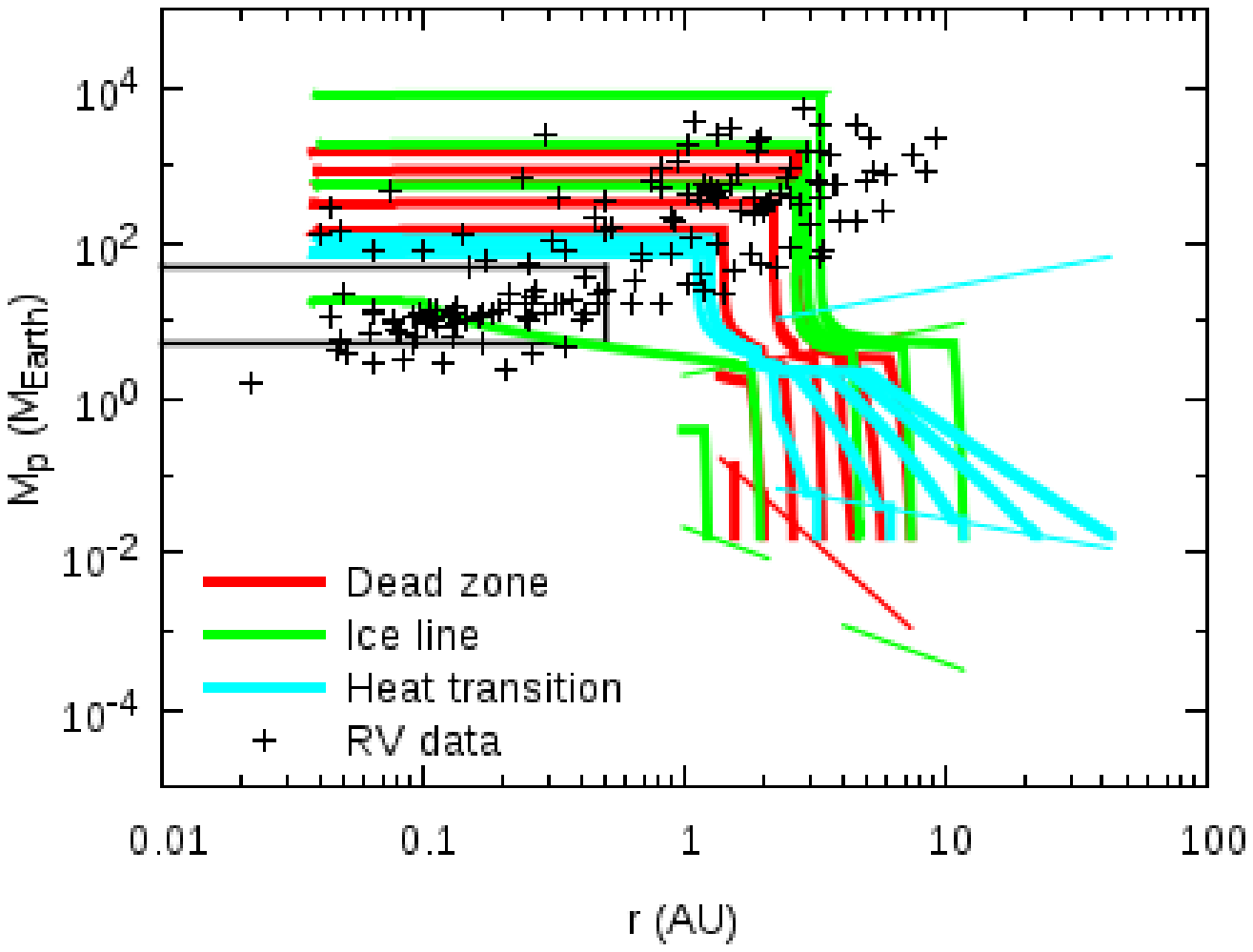}
\includegraphics[width=8cm]{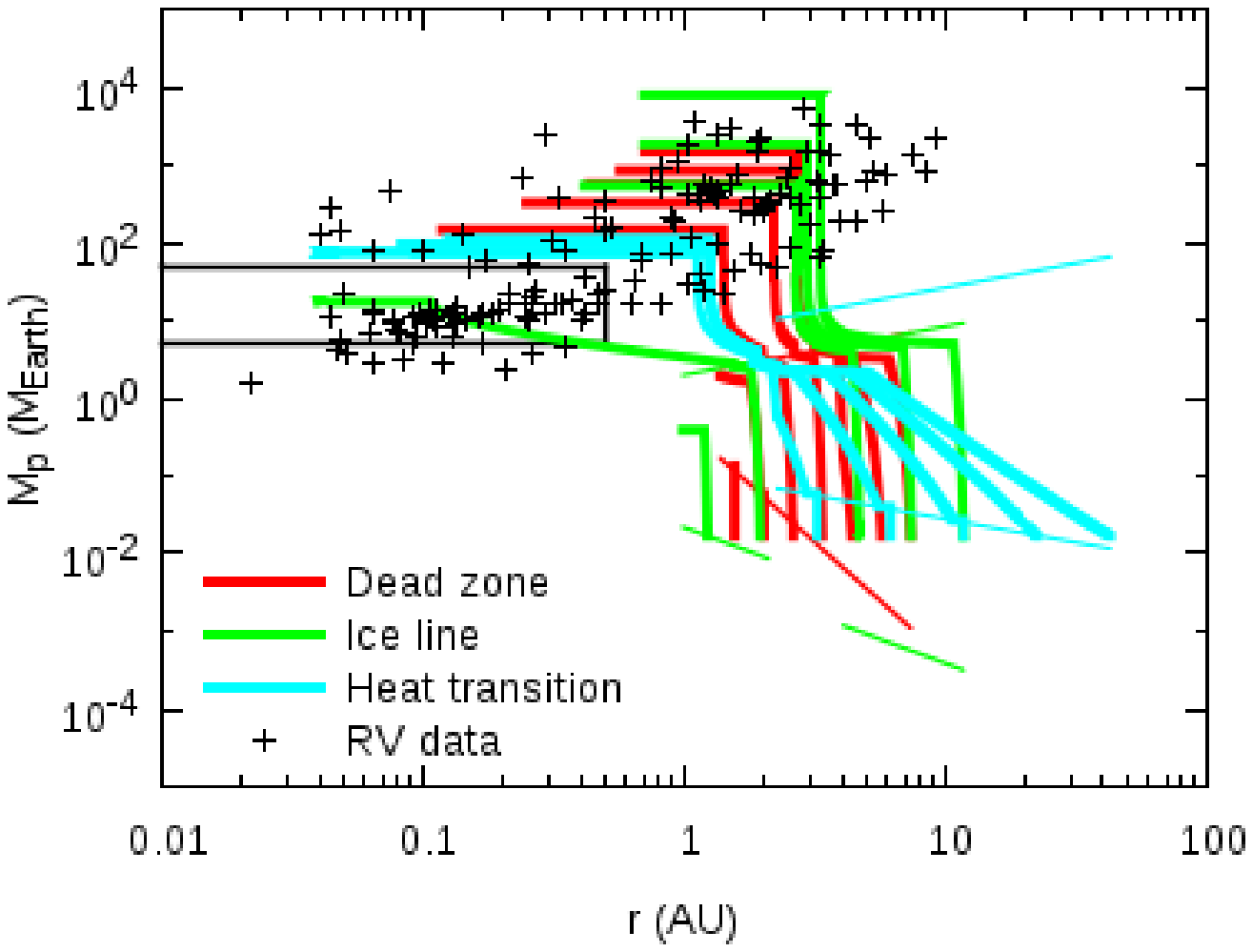}
\includegraphics[width=8cm]{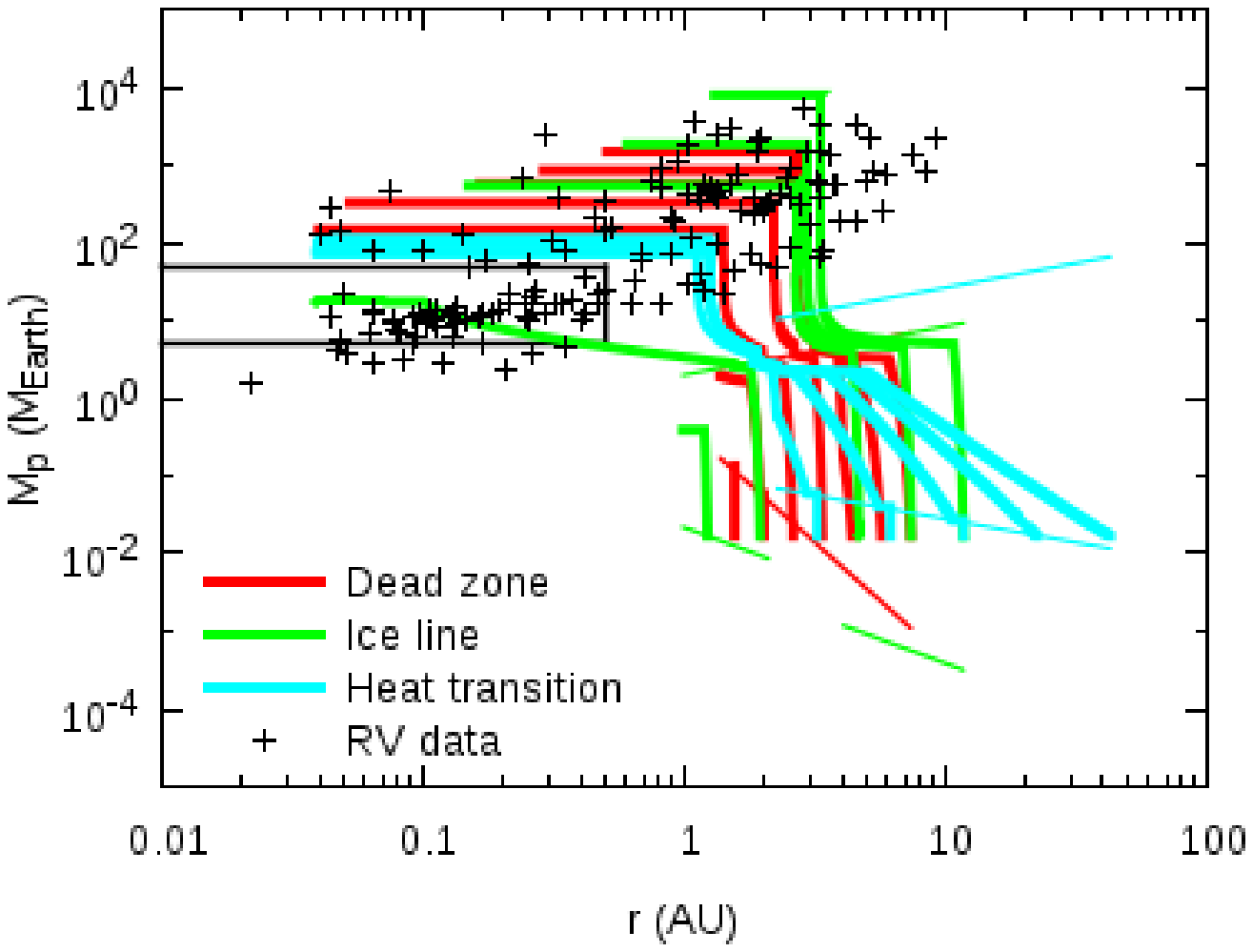}
\includegraphics[width=8cm]{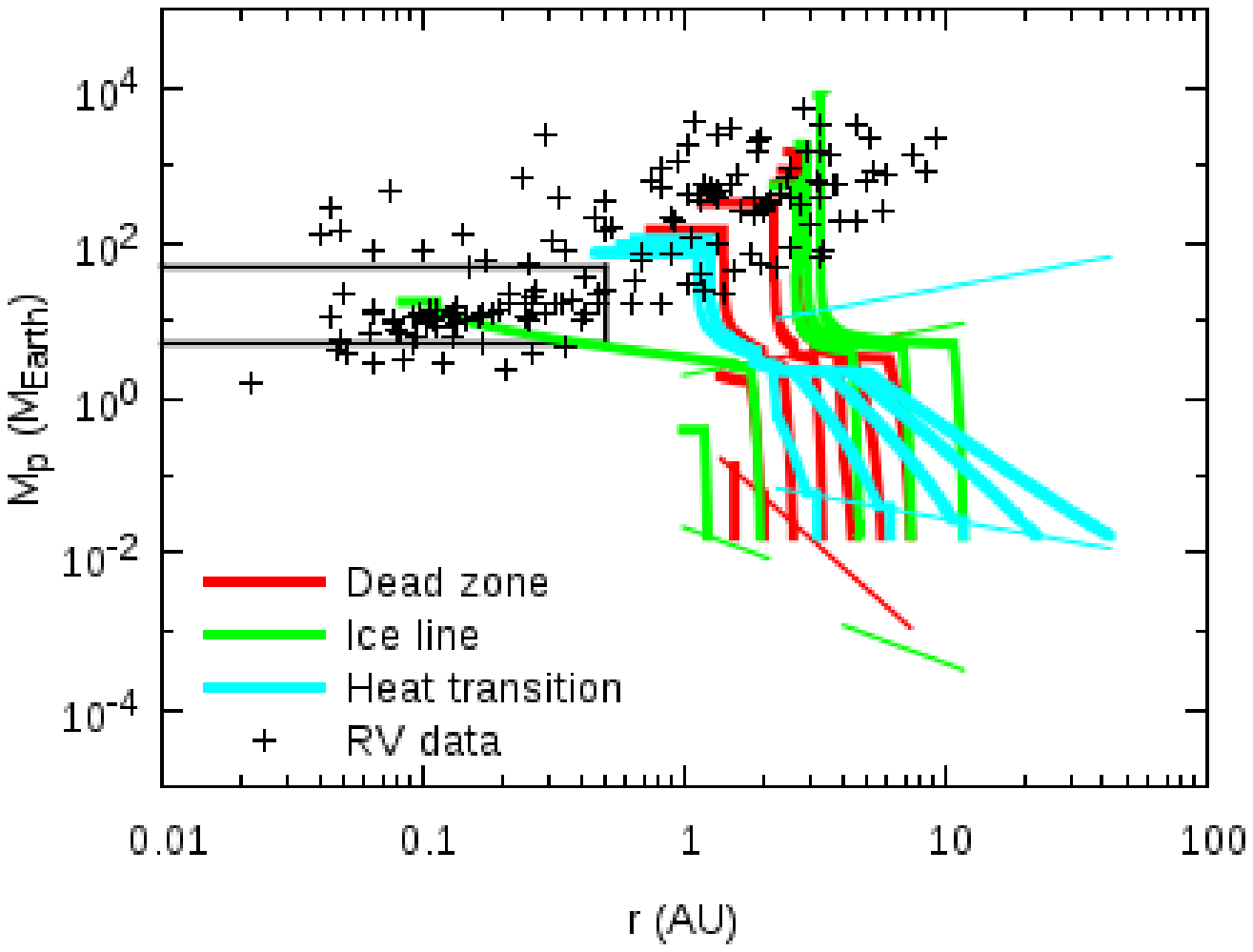}
\caption{Parameter study of the effects of slower type II migration (see Table \ref{table7}). The resultant 
evolutionary tracks of planets that grow in all the three planet traps are shown (as Fig. \ref{fig5}). The top left panel denotes the results of 
Run B1, the top right for Run B2, the bottom left for Run B3, and the bottom right for Run B4. For two top panels, 
the slowing rate is independent of planetary mass while it depends for two bottom panels. When no slowing 
mechanism is in action (Run B1, $f_{mig,slowerII}=0$), the observed population of exoplanets cannot be reproduced 
even if planetary cores are captured and distribute by moving planet traps. The observations can be reproduced only if 
the standard type II migration sufficiently slows down ($f_{mig,slowerII}\gtrsim 1$ or $f_{mig,slowerII} = \bar{f} M_{crit}/M_{p}$ with 
$\bar{f} \gtrsim 1$). It is important that mass dependency on the slowing mechanism of the type II migration does not matter. Thus, 
this suggests that planet traps are more crucial for understanding the statistics of observed exoplanets.}
\label{fig9}
\end{center}
%\end{figure}
\end{minipage}
\end{figure*}

Fig. \ref{fig9} shows the results of the evolutionary tracks of planets growing in all the three planet traps. The top 
left panel shows the results of Run B1, the top right for Run B2, the bottom left for Run B3, and the bottom right for 
Run B4. For two top panels, there is no mass dependency in equation (\ref{slower_typeII}) while it depends on 
planetary mass for two bottom panels. Comparison of Run B1 (Top left) with any other runs demonstrates that some 
kind of process which slows down the standard type II migration is clearly needed for reproducing the observed population 
of exoplanets even if planetary cores are saved due to planet traps. This is because the standard type II migration that 
proceeds as local viscous timescale leads planets to spiraling into the host stars within the disk lifetime at $r\lesssim$ 10 AU. 

When the type II migration somewhat slows down sufficiently, our findings discussed above, especially the mass-period 
relation, are valid for various cases (see Top right and two Bottom panels). It is important that the overall feature 
of the resultant populations is very insensitive to the origins of slowing down mechanisms (mass dependence vs 
independence, compare top right panel with Fig. \ref{fig5}), although the radial distribution somewhat varies for each case. 
We confirmed, through experiments, that the observed planetary population can be reproduced when 
$f_{mig,slowerII}\gtrsim 1$ or $f_{mig,slowerII} = \bar{f} M_{crit}/M_{p}$ with $\bar{f} \gtrsim 1$. 

In addition to the slower type II migration, photoevaporation of gas is also important for establishing the final 
radial distribution of planets, since it terminates the migration. The combined effects of type II migration and 
photoevaporation on planetary populations were discussed in the literature \citep[e.g.][]{all02,mjm03,a07,aa09,ap12}. 
Our models are more fundamental than theirs in a sense that we have incorporated the effects of type I migration which 
can be captured at planet traps. We will investigate in detail the combined effects in the forthcoming paper.  

In summary, we can conclude that planet traps, not slower type II migration, play the primary role in reproducing 
the observations, provided that the standard type II migration slows down sufficiently when planetary mass exceeds the 
local disk mass.

\begin{table}
\begin{center}
\caption{Parameter study of slower type II migration}
\label{table7}
\begin{tabular}{cc}
\hline
        &  $f_{mig,slowerII}$       \\ \hline
Run B1  &  0                        \\
Run B2  &  1 $\times M_{crit}/M_p$ \\
Run B3  &  0.1                      \\
Run B4  &  10                       \\
\hline
\end{tabular}
\end{center}
\end{table}

\section{Discussion \& Conclusions} \label{conc}

We have constructed and followed the evolutionary tracks of protoplanets generated by combining core 
accretion together with the movement of planet traps in evolving disks. We have focused on three types of 
inhomogeneities in protoplanetary disks and the resultant planet traps: dead zones, ice lines, and heat transitions. 

We have demonstrated that the plant traps play two fundamental roles in planet formation and migration. The first 
is to trap cores of gas giants that otherwise undergo rapid type I migration. This trapping leads to the formation 
of gas giants orbiting at 0.01 AU $\lesssim r \lesssim$ 10 AU without the cores falling into the central star before 
disk evolution is terminated by disk photoevaporation. The second is to transport the trapped cores very slowly 
from large to small periods. This transport distance is regulated by the gap opening mass and the properties of the 
planet traps and hence is well coupled with planetary growth histories; different planet traps result in different 
efficiencies of planet formation and migration. Consequently, planet traps are regarded effectively as a filter for 
distributing cores - massive cores tend to hover around large periods while low-mass cores around short periods. 

We have seen that the combination of planet traps, planetary growth and type II migration in evolving disks generates 
planetary populations that have a wide range in mass and period. The final positions of planets are determined when 
photoevaporation of gas disks (rather than disk viscosity) plays the dominant role in disk evolution. 
It is noted that the final distribution of planets can be largely affected by the combination of slower type II migration 
and the photoevaporation rate that is adjusted by $f_{pe}$ (see equation (\ref{mdot_pe}) and Table \ref{table2}). 
In order to examine this dependency, we will perform a more comprehensive parameter study in a subsequent paper. 
In addition, different accretion histories of planets will result in differences in the composition of the planets and their 
atmospheres, which can be investigated by extending our models.

What about purely dynamical effects arising from planet-planet interactions? These are well known to be important for 
understanding the observed eccentricity distribution \citep[e.g.][]{rf96}. However, \citet{mtc10} have recently 
clarified through numerical simulations that the semi-major axes of planets are determined mainly by planetary 
migration in the gas disks while the eccentricities are determined after the gas disks dissipate severely. This 
indicates that the results of planetary evolution in the gas disk phase will not be washed out by subsequent planetary 
dynamics.    

We list our major findings below.
 
\begin{enumerate}

\item We have demonstrated that the wide range of end points of planets evolving in the mass-semi-major axis diagram 
establishes a theoretical mass-period relation in which planetary mass is an increasing function of orbital period 
(see Fig. \ref{fig5}). This is in excellent agreement with the observational data in a sense that the data scatter 
around the end point of our evolutionary tracks.

\item We have shown that the many tracks of dead zones and ice lines preferentially tend to end up at $\sim$ 1 AU (see 
Fig. \ref{fig5}). Combined with an argument that planet formation efficiencies are reasonably high there, the preference 
provide a physical explanation for the pile up of observed gas giants at $\sim$ 1 AU.  

\item We have also demonstrated that planets that grow in dead zone traps end up at $r \sim$ 1 AU, ice line traps at 
0.03AU $\lesssim r \lesssim$ 3 AU, and heat transition traps at $r\sim$ 0.1 AU (see Fig. \ref{fig4}). The resulting 
wide range of planets in the mass-semi-major axis diagram is insensitive to the detailed structure of dead zones 
and accounts for a number of important observational trends.  

\item We have also shown that moving ice line traps can put planets in the planet deserts that were predicted by the 
earlier population synthesis models (see Fig. \ref{fig5}). As denoted by Fig. \ref{fig1}, the desert is located in 
the range of planetary masses (5-50 $M_{\oplus}$) and semi-major axes (0.04-0.5 AU). The recent observations discover 
many planets in the deserts. Thus, our models are more consistent with the observations in this regard. 

\item We predict planet deserts that arise from the nature of planet traps. They have physically different 
origins from the earlier claimed ones. Our deserts are relevant only when protoplanetary disks have sufficient amount 
of gas that drives rapid type I migration. This suggests that the deserts can be filled by planets that form far beyond 
the deserts and eventually migrate there in gas disks. The planets are likely to emerge after the gas disks 
severely disperse and/or to be transported there due to planet-planet scatterings. Our deserts are present in the range of 
planetary masses (1-50 $M_{\oplus}$) and semi-major axes (1-10 AU), which covers the primary target of ongoing and 
future observational surveys such as the HARPS and Kepler missions.

\item The more massive the host star, the more the evolutionary tracks in the mass-period diagram are pushed towards 
large disk radii. This arises because of the much more rapid accretion rates in more massive systems 
($\dot{M} \propto M_*^2$).

\end{enumerate}

In the forthcoming paper, we will use N-body simulations to take into account the physics of planet-planet interactions 
that can be induced by growing planets in different planet traps.

%% The \notetoeditor{TEXT} command allows the author to communicate
%% information to the copy editor.  This information will appear as a
%% footnote on the printed copy for the manuscript style file.  Nothing will
%% appear on the printed copy if the preprint or
%% preprint2 style files are used.

\acknowledgments
The authors thank Kees Dullemond, Shigeru Ida, Hubert Klahr, Soko Matsumura, Chris McKee, Christoph Mordasini, 
Takayuki Muto and Taku Takeuchi for stimulating discussions, and an anonymous referee for useful comments on our 
manuscript. Also, YH thank the hospitality of ITA, University of 
Heidelberg and Tokyo Institute of Technology for hosting stimulating visits. YH is supported by McMaster 
University, as well as by Graduate Scholarships from the ministry of Ontario (OGS) and the Canadian Astrobiology 
Training Program (CATP).  REP is supported by a Discovery Grant from the Natural Sciences and Engineering Research 
Council (NSERC) of Canada.




%\clearpage

\appendix

\section{A: Characteristic masses in the mass-semi-major axis diagram} \label{app1}

\begin{figure*}
\begin{minipage}{17cm}
%\begin{figure}%[!ht]
\begin{center}
\includegraphics[width=8cm]{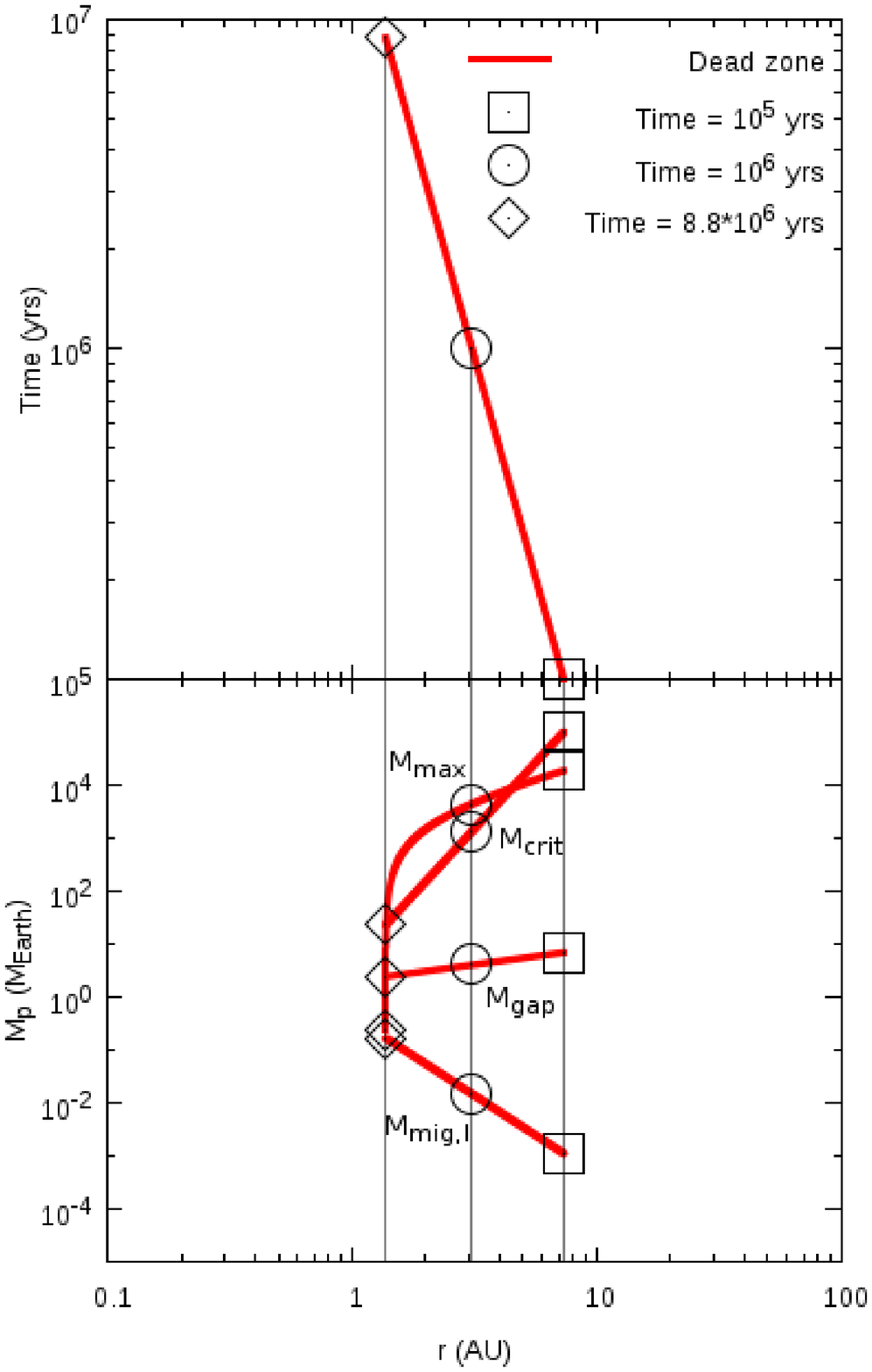}
\includegraphics[width=8cm]{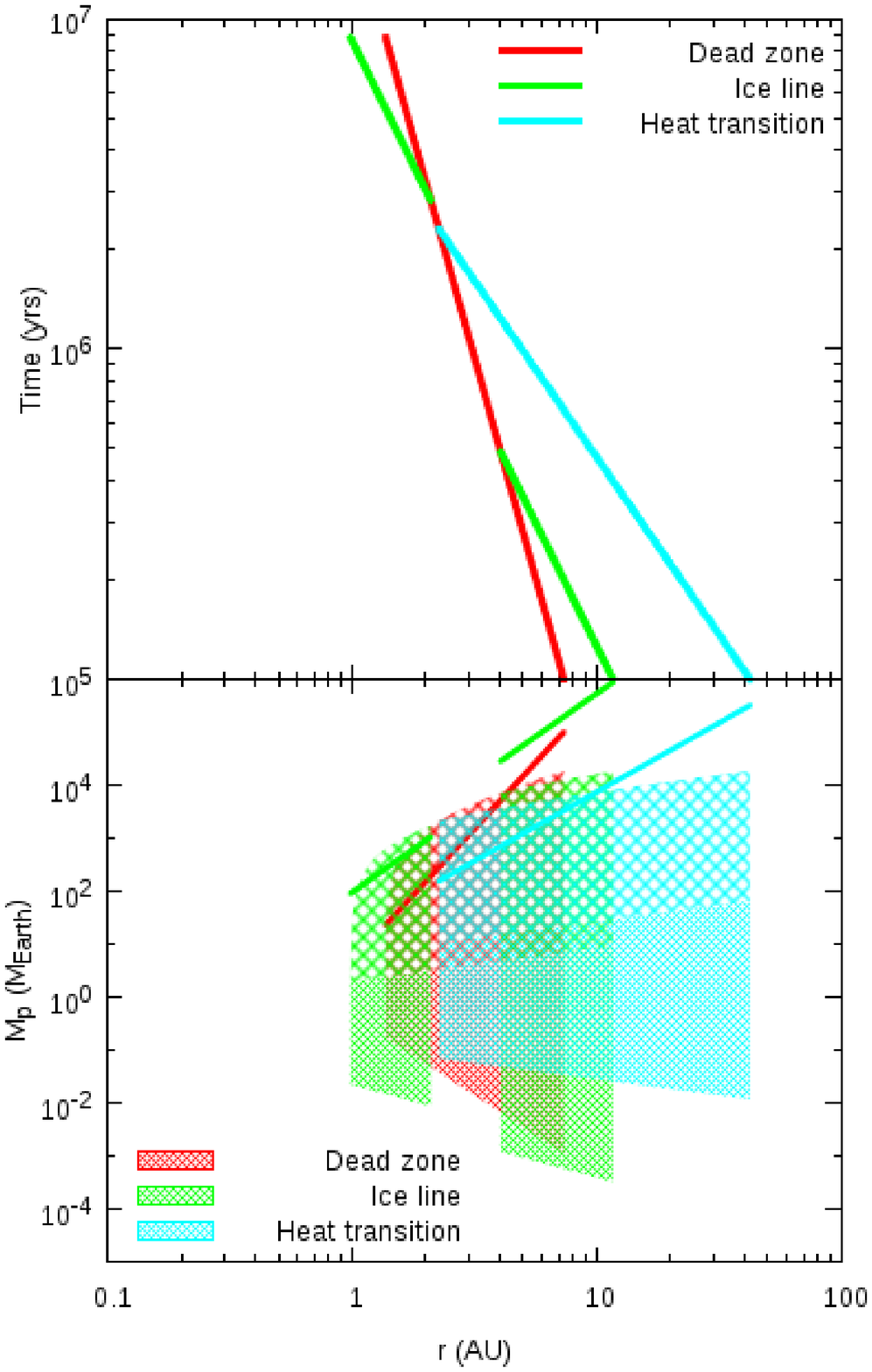}
\caption{{\it Left}: (Top) The movement of a dead zone trap in evolving disks. The initial position is determined by 
assuming a start time of $10^{5}$ years. The final position arises when the accretion rate $\dot{M}$ equals the 
photoevaporation rate $\dot{M}_{pe}$, - a condition which also defines the disk lifetime 
($\tau_{disk} \simeq 8.8\times 10^{6}$ years in our fiducial model). (Bottom) The evolution of four characteristic 
masses at the dead zone trap. The top line represents $M_{max}$, the second for $M_{crit}$, the third for $M_{gap}$, 
and the lowest one for $M_{mig,I}$ (see Table \ref{table3}). Every position of the dead zone (shown in the top panel) 
defines four masses (see some symbols as examples), and as the dead zone moves inward, they also move inward. 
{\it Right}: (Top) The movement of all three traps in evolving disks. The dead zone trap is denoted by red, the ice line 
by green, and the heat transition by light blue. They move inward due to viscous evolution of the disks, but at different 
rates. (Bottom) The evolution of four characteristic masses at all three traps. The color scheme is the same as the top 
panel. The type II regimes that are defined by $M_{max}$ and $M_{gap}$, are denoted by the coarse hatched regions while 
the trapping regimes that are specified by $M_{gap}$ and $M_{mig,I}$ are represented by the fine hatched regions. The 
solid lines denote $M_{crit}$. The disappearance and re-emergence of these regimes correspond to those of the positions 
of the planet traps that are shown in the top panel.}
\label{figA}
\end{center}
%\end{figure}
\end{minipage}
\end{figure*}

We discuss the segmentation of a diagram for planetary mass versa semi-major axis. As an example,the bottom panel of 
Fig. \ref{figA} (Left) shows evolution of four characteristic masses ($M_{max}$, $M_{crit}$, $M_{gap}$, and $M_{mig,I}$) 
at a dead zone for the fiducial case (also see Table \ref{table3}). Every position of the dead zone that is specified by 
the time $\tau$ defines four masses (see circles on Fig. \ref{figA} (Left) as an example). As the dead zone moves 
inwards following the disk evolution (see the top panel of Fig. \ref{figA} (Left)), these four masses also move inwards 
at the same rate. As a result, four lines are drawn that track the evolution of the characteristic masses in the 
diagram. The top line denotes $M_{max}$, the second for $M_{crit}$, the third for $M_{gap}$, and the bottom for 
$M_{mig,I}$. Thus, the top and third lines define the boundaries of type II migration in the mass-semi major axis 
diagram while the third and bottom one defines the boundaries for trapping type I migration. The second line defines 
where the inertia of planetary mass becomes effective that can slow down type II migration.

The bottom panel of Fig. \ref{figA} (Right) shows the evolution of four characteristic masses at all three disk 
inhomogeneities for the fiducial case. The type II regimes for each inhomogeneity are denoted by the coarse hatch 
regions while the fine hatch ones represent the type I trap regimes. The solid lines denote $M_{crit}$. The dead zone 
is denoted by red, the ice line by green, and the heat transition by light blue. Disappearance and re-appearance of these 
regions correspond to the behavior of each inhomogeneity that is shown in the top panel.

%\appendix

\section{B: Analytical prescriptions for planetary growth} \label{app2}

We summarize the standard results of the core accretion scenario needed to follow the growth of planets as they move in 
disks.

\subsection{Stage I: Formation of cores}

Formation of cores in planetesimal disks is well understood in the literature \citep[e.g.][]{ki02}, and the growth 
timescale in the disks is given as
\begin{eqnarray}
 \tau_{c,acc} & \simeq & 1.2 \times 10^{5} \mbox{ yr} \left( \frac{\Sigma_d}{10 \mbox{ g cm}^{-2}} \right)^{-1}
                    \left( \frac{r}{r_0} \right)^{1/2} \left( \frac{M_c}{M_{\oplus}} \right)^{1/3}
                    \left( \frac{M_*}{M_{\odot}} \right)^{-1/6} \\ \nonumber
             & \times & \left[ \left( \frac{b}{10} \right)^{-1/5}
                        \left( \frac{\Sigma_g}{2.4 \times 10^3 \mbox{ g cm}^{-3}} \right)^{-1/5} 
                       \left( \frac{r}{r_0} \right)^{1/20} \left( \frac{m}{10^{18} \mbox{ g}} \right)^{1/15}
                     \right]^2,
 \label{tau_cacc}
\end{eqnarray}
where $\Sigma_d$ is the surface density of dust, $M_c$ is the mass of a core, $b=10$ is a parameter for determining 
the feeding zone of the core (see below), and $m=10^{18}$g is the mass of planetesimals that are accreted onto the core. 
Adopting this timescale, the growth of cores is regulated by 
\begin{equation}
 \frac{d M_p}{dt} =\frac{M_p}{\tau_{c,acc}}.
 \label{growth1}
\end{equation}

\subsection{Stage II: slow gas accretion onto the envelopes}

As cores grow, their feeding zones $\triangle r_c$ empty. These zones are scaled by $b r_H$, where 
$r_H=(M_c/(3M_*))^{1/3}$ is the Hill radius of a core and $b \sim 10$. The decrease of planetesimals in the zones 
results in the reduction of the accretion rate of cores $\dot{M}_c$. In the limit of the modest to high velocity 
dispersion $\sigma$ of planetesimals that are accreted onto the cores, $\dot{M}_c$ can be written as \citep{s72,il04i}
\begin{equation}
 \dot{M}_c \sim 2 \pi \left( \frac{R_c}{r} \right) \left( \frac{M_c}{M_*} \right) 
                   \left( \frac{r\Omega}{\sigma}\right)^{2} \Sigma_d r^2 \Omega, 
\end{equation}
where $R_c$ is the radius of cores. Planetesimals within the feeding zones can reach cores when 
$\sigma/\Omega \sim \triangle r_c (=b r_H)$. Assuming $R_c$ to be similar to that of the Earth;  
\begin{equation}
 R_c = 6.4 \times 10^{8} \mbox{ cm} \left( \frac{M_c}{M_{\oplus}}\right)^{1/3} 
                                   \left( \frac{\rho_c}{5.5 \mbox{ g cm}^{-3}} \right)^{-1/3},
\end{equation}
$\dot{M}_c$ is given as
\begin{equation}
   \dot{M}_c  \sim  3.0\times 10^{-8} M_{\oplus} \mbox{ yr}^{-1} \left( \frac{b}{10} \right)^{-2} 
                 \left( \frac{\rho_c}{5.5 \mbox{ g cm}^{-3}} \right)^{-1/3} 
             \left( \frac{M_c}{M_{\oplus}} \right)^{2/3} \left( \frac{M_*}{M_{\odot}} \right)^{-1/3} 
                   \left( \frac{\Sigma_d}{10 \mbox{ g cm}^{-2}} \right)
                 \left( \frac{r}{r_0} \right) \left( \frac{\mbox{ yr}}{1 /\Omega} \right).
\end{equation}

When all the planetesimals in their feeding zones are consumed, the cores attain the maximum mass that is known as the 
isolation mass, which is defined by \citep{ki02,il04i} 
\begin{equation}
 M_{c,iso} =  2 \pi r \vartriangle r_c \Sigma_d 
            \simeq  0.16 M_{\oplus} \left( \frac{b}{10} \right)^{3/2} 
                      \left( \frac{\Sigma_d}{10 \mbox{ g cm}^{-2}} \right)^{3/2} \left( \frac{r}{r_0} \right)^{3}
                      \left( \frac{M_*}{M_{\odot}} \right)^{-1/2}. 
\end{equation}
The accretion of gas onto cores and subsequent envelope formation are initiated when the mass of core becomes larger 
than 
\begin{equation}
 M_{c,crit} \simeq 10 f_{c,crit} M_{\oplus} \left( \frac{\dot{M}_c}{10^{-6}M_{\oplus} \mbox{ yr}^{-1}} \right)^{1/4}.
 \label{m_ccrit}
\end{equation}
This critical mass was originally derived from a series of numerical simulations that investigated the effect of cores' 
accretion rates and opacity in the envelope on formation of gas giants \citep{ine00}. Here, we have adopted a simplified 
one, following \citet{il04i}. Recent studies, however, revealed that $M_{c,crit}$ might be smaller than that predicted 
by equation (\ref{m_ccrit}) with $f_{c,crit}=1$ \citep[e.g.][]{hi11}. In order to take this into account, we have 
introduced a dimensionless factor $f_{c,crit}$. 

The gas accretion rate of cores is prescribed by \citep{il04i}
\begin{equation}
 \frac{d M_p}{dt} \simeq \frac{M_p}{\tau_{KH}},
 \label{growth2}
\end{equation}
where the Kelvin-Helmholtz timescale is given as
\begin{equation}
 \tau_{KH} \simeq 10^{c} \mbox{ yr} \left( \frac{M_p}{M_{\oplus}} \right)^{-d},
 \label{tau_KH}
\end{equation}
where $c=8$ and $d=2.5$. This is a simplified timescale that was originally estimated from numerical simulations 
\citep{ine00}. As shown by the more detailed numerical simulations \citep{p96,lhdb09}, this stage is slow 
($\gtrsim 10^{6}$ years). 

\subsection{Stage III: runaway gas accretion onto the cores}

When planets become massive enough, runaway gas accretion onto the cores starts. In the detailed numerical simulations, 
this stage commences when the envelope of cores becomes more massive than the cores \citep{p96,ine00,lhdb09}. 
Nonetheless, we monitor this stage by the condition that 
\begin{equation}
 \frac{\tau_{KH}}{10^5 \mbox{ yr}} < 1.
\end{equation}
This is because our approach is rather simple. This condition never affects our results. The growth rate of this stage 
is also prescribed by equation (\ref{growth2}). 

It is totally unclear how gas accretion onto the cores terminates and what physical process(es) determines the final 
mass of gas giants. Therefore, we assume that Stage III continues until planets gain the mass $f_{max}M_{max}$, where 
$f_{max}$ is an adjustable parameter. 
 
\subsection{Parameters for planetary growth} \label{para_growth}

As discussed above, planetary growth and consequent evolutionary tracks of planets are regulated by four parameters in 
our model (see Table \ref{tableB}). The parameter $f_{c,crit}$ governs the onset of gas accretion of cores, a set of 
parameters $c$ and $d$ determine the efficiency of gas accretion onto cores, and $f_{max}$ controls the final mass of 
planets. We denote these values given in Table \ref{tableB} our fiducial model. Note that \citet{il04i} adopted the 
values of $f_{c,crit}=1$, $c=9$, and $d=3$ rather than our set of $f_{c,crit}=0.3$, $c=8$, and $d=2.5$. We performed a 
parameter study and confirmed that different choice of these values does not change our results very much. Therefore, 
our choice and results are robust in a sense that we can compare our results with the results of \citet{il04i,il08v}. 
We present only a parameter study of $f_{max}$ in Appendix \ref{app3}, because the choice of this value is probably 
the most uncertain. 

%\begin{table*}
%\begin{minipage}{15cm}
\begin{table}
\begin{center}
\caption{Important parameters for planetary growth}
\label{tableB}
\begin{tabular}{ccc}
\hline
Symbols       &  Meaning                                                                     & Value    \\ \hline
$f_{c,crit}$  &  A factor linked to the critical mass of cores (equation (\ref{m_ccrit}))    & 0.3      \\
$c$           &  Exponent of the Kelvin-Helmholtz timescale (equation (\ref{tau_KH}))        & 8        \\
$d$           &  Exponent of the Kelvin-Helmholtz timescale (equation (\ref{tau_KH}))        & 2.5      \\
$f_{max}$     &  A factor linked to the maximum mass of planets                              & 0.1      \\
\hline
\end{tabular}
\end{center}
\end{table}
%\end{minipage}
%\end{table*}

%\appendix

\section{C: A parameter study for planet growth} \label{app3}

As discussed in Appendix \ref{app2}, our choice of most parameters that regulate planetary growth is based on the 
physical considerations and the more recent results of detailed simulations. Hence our choice is compatible with the 
original work of \citep{il04i}. However, there is one exception, which is $f_{max}$ which constrains the maximum mass 
of planets. This is mainly because there are no firm physical arguments and simulations of how the final mass of planets 
is established. Therefore, we now examine how the value of $f_{max}$ affects our results.

\begin{figure*}
\begin{minipage}{17cm}
%\begin{figure}%[!ht]
\begin{center}
\includegraphics[width=8cm]{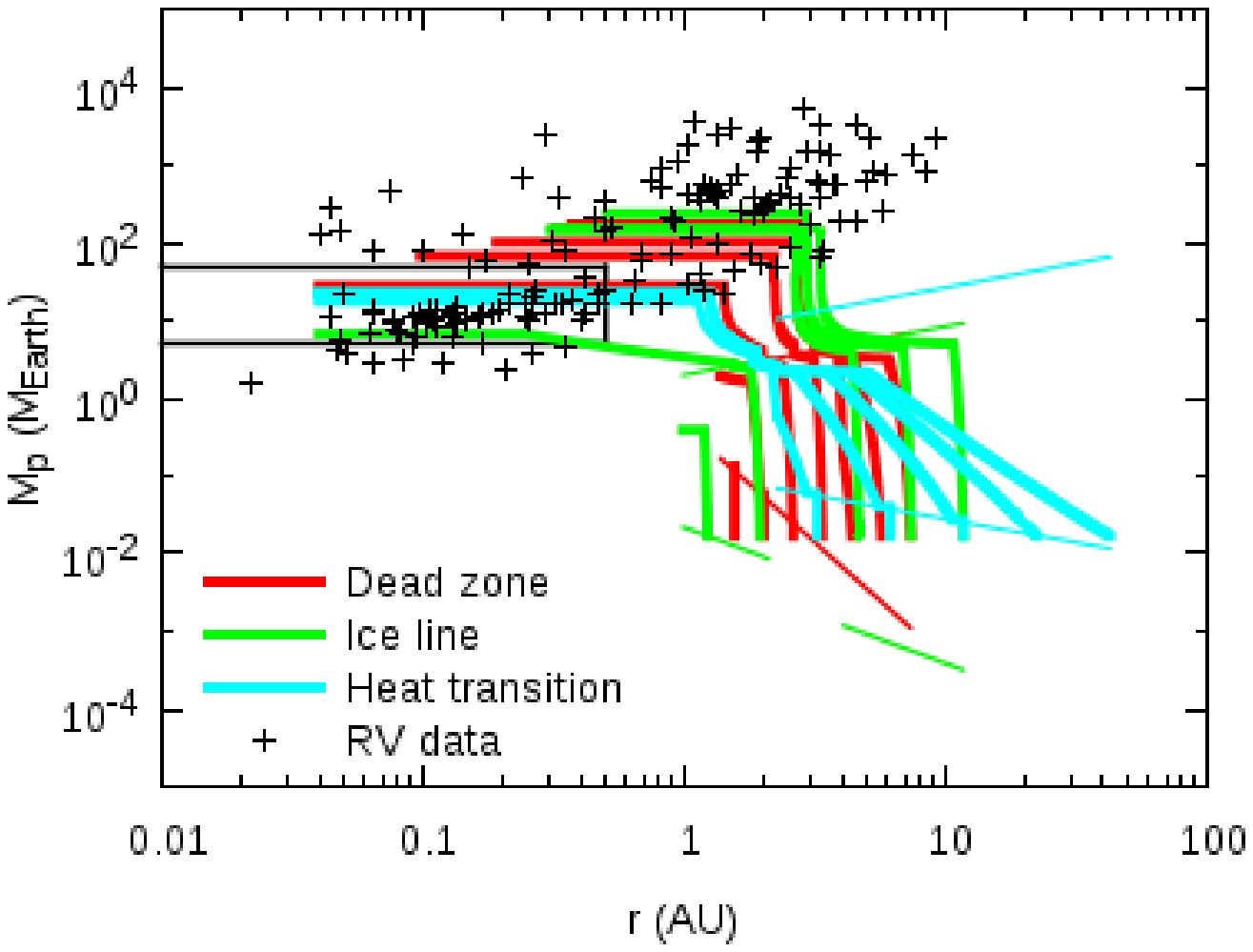}
\includegraphics[width=8cm]{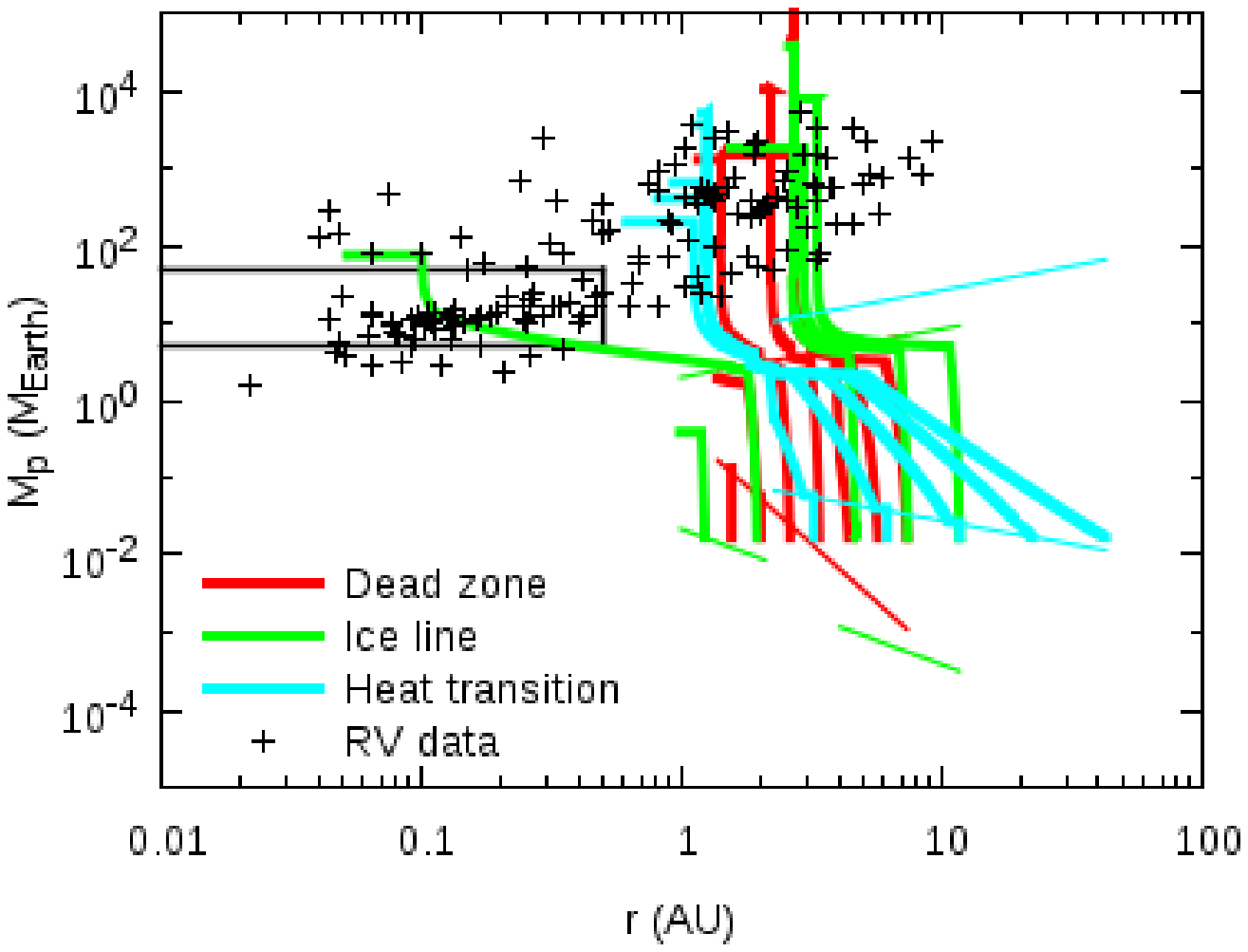}
\caption{Parameter study of the variation of maximum mass of planets $f_{max}$ (see Table \ref{tableC}). The 
evolutionary tracks of planets that grow in all three planet traps are shown (as Fig. \ref{fig5}). The result of Run C1 
($f_{max}=0.03$; low mass case) is presented in the left panel while that of Run C2 ($f_{max}=0.3$; high mass case) in 
the right. Although the pile up at 1 AU and low mass planets with small orbital radii are slightly affected by 
$f_{max}$, the resultant populations are still well understood by the physical considerations discussed in $\S$ 
\ref{resu}.}
\label{figC}
\end{center}
%\end{figure}
\end{minipage}
\end{figure*}

Table \ref{tableC} summarizes our parameter study on $f_{max}$. For Run C1, we took $f_{max}=0.03$ while $f_{max}=0.3$ 
for Run C2. Except for the value of $f_{max}$, we adopted the same values of the fiducial model. Fig. \ref{figC} shows 
the results of the evolutionary tracks of planets for both cases. The left panel shows the results of Run C1 while the 
right one for the Run C2. The results of both cases are generally very similar to that of the fiducial model. More 
specifically, both cases produced theoretical mass-period relations that are broadly consistent with the observations. 
On closer examination, we see that the pile up of gas giants at $\sim 1$ AU and the presence of low-mass planets with 
small orbital radii are relatively affected by the value of $f_{max}$. This is indeed expected. If $f_{max}$ has a 
small value, then low mass planets become the main product and planet formation completes earlier. Consequently, the 
planets experience substantial inward type II migration. In addition, the effect of the inertia of the planets that 
slows down the type II migration is also reduced. This results in larger populations of low-mass planets at small 
orbital radii. Thus, models with small values of $f_{max}$ have difficulty in reproducing the pile up at 1 AU. The 
opposite happens for large values of $f_{max}$. In this case, massive planets are preferentially formed and the 
completion of planet formation takes place at later time, so that the subsequent inward type II migration is 
significantly suppressed due to both the shorter remaining time and the larger inertia of planets. As a result, 
the population of low-mass planets with tight orbits declines while the 1 AU pile up is enhanced. 

In summary, the results depend slightly on some basic parameters such as $f_{max}$. Nonetheless, they are well 
understood by the physical arguments presented in $\S$ \ref{resu}. Therefore, our findings are reasonably robust 
for a wide range of the parameter space.   

\begin{table}
\begin{center}
\caption{Parameter study of planetary growth}
\label{tableC}
\begin{tabular}{cc}
\hline
        &  $f_{max}$  \\ \hline
Run C1  &  0.03       \\
Run C2  &  0.3        \\
\hline
\end{tabular}
\end{center}
\end{table}

\bibliographystyle{apj}

\bibliography{apj-jour,adsbibliography}    %% includes the journal abbrevs

\end{document}